%% file: main.tex
\documentclass[
 reprint,
 nofootinbib,
 amsmath,amssymb,
 aps,
 twocolumn,
 notitlepage
]{revtex4-1}

\usepackage{braket}
\usepackage[normalem]{ulem}
\usepackage[final]{microtype}
\usepackage{flushend}
\usepackage{algorithm}
\usepackage{algpseudocode}
\usepackage{soul}
\usepackage{styles}
\usepackage{graphicx}
\usepackage{subcaption}
\usepackage{mathtools}
\usepackage{caption}
\usepackage{multirow}
\usepackage{xcolor}     

\DeclarePairedDelimiter{\ceil}{\lceil}{\rceil}

\algblockdefx[Name]{Do}{End}%
[2][Unknown]{\textbf{do Parallel }}%
   {\textbf{end do}}

\begin{document}

\title{Impact of qubit connectivity on quantum algorithm performance}

\author{Adam Holmes}
\affiliation{University of Chicago}
\affiliation{Intel Corporation}
\author{Sonika Johri}
 \email{sonika.johri@intel.com}
\affiliation{Intel Corporation}
\author{Gian Giacomo Guerreschi}
\affiliation{Intel Corporation}
\author{James S. Clarke}
\affiliation{Intel Corporation}
\author{A. Y. Matsuura}
\affiliation{Intel Corporation}


\date{\today}


\begin{abstract}
Quantum computing hardware is undergoing rapid development from proof-of-principle devices to scalable machines that could eventually challenge classical supercomputers on specific tasks. On platforms with local connectivity, the transition from one- to two-dimensional arrays of qubits is seen as a natural technological step to 
increase the density of computing power and to reduce the routing cost of limited connectivity. Here we map and schedule representative algorithmic workloads - the Quantum Fourier Transform (QFT) relevant to factoring, the Grover diffusion operator relevant to quantum search, and Jordan-Wigner parity rotations relevant to simulations of quantum chemistry and materials science - to qubit arrays with varying connectivity. In particular we investigate the impact of restricting the ideal all-to-all connectivity to a square grid, a ladder and a linear array of qubits. Our schedule for the QFT on a ladder results in running time close to that of a system with  all-to-all connectivity. Our results suggest that some common quantum algorithm primitives can be optimized to have execution times on systems with limited connectivities, such as a ladder and linear array, that are competitive with systems that have all-to-all connectivity.”


\vspace{1.5cm}
\end{abstract}

\maketitle





\section{Introduction}
\label{sec:introduction}
\input{introduction.tex}

\section{Summary of Results}
\label{sec:summary}
\input{summary.tex}

\section{Background and notation}
\label{sec:background}
\input{background.tex}

\section{Hardware architectures}
\label{sec:hardware}
\input{hardware.tex}

\section{Representative Algorithmic Workloads}
\label{sec:algorithms}
\input{algorithms.tex}

\section{Results}
\label{sec:results}
\input{results.tex}
\section{Conclusion and Discussion}
\label{sec:conclusion}
\input{conclusion.tex}


\newpage
\onecolumngrid

\bigskip\noindent\makebox[\linewidth]{\resizebox{0.54\linewidth}{1pt}{$\bullet$}}\bigskip

\appendix
\renewcommand\thefigure{\thesection.\arabic{figure}}
\renewcommand\thetable{\thesection.\arabic{table}}
\setcounter{figure}{0}
\setcounter{table}{0}
\vspace{1cm}
\section{Gate Decompositions}
\input{gate-decompositions.tex}
\label{sec:gate-decompositions}

\newpage


%

\end{document}

%% file: introduction.tex

Quantum devices may one day be able to perform computational tasks that exceed the capabilities of classical computers. To this end, quantum algorithms as well as quantum hardware are active areas of research. Given the challenging nature of this goal, co-design between quantum hardware and quantum algorithms will be integral to constructing useful quantum computers. In this paper we present research addressing how the physical qubit layout affects its suitability for a specific application area.

Quantum algorithms are nowadays developed using high-level abstractions in which quantum circuits are described in terms  of an ideal qubit register in which two- or multi-qubit gates can be performed between any subset of qubits. In practice though, almost all physical architectures will allow for two-qubit operations between a limited set of qubit pairs, a property that can be termed as the connectivity of the qubits. Additional routing operations or teleportation protocols are then required to implement the desired two-qubit gates, and these solutions may introduce a substantial overhead in terms of the execution time of the algorithm or in the number of ancilla qubits. This makes highly-connected devices, where two-qubit gates can be performed between as many pairs of qubits as possible, seemingly favorable platforms. However, higher connectivity devices are harder to manufacture and the difficulty is experienced across multiple technologies and architectures ranging from ion traps \cite{lekitsch2017blueprint}, to superconducting qubits \cite{barends2014superconducting,versluis2017scalable}, to quantum dots \cite{li2017crossbar}.
The impact of technological limitations can be reduced by developing compilation tools and scheduling techniques that soon will play an essential role in the operation of real quantum computers. Since a majority of quantum algorithms leverage a relatively small set of algorithmic primitives, optimizing the scheduling of these to physically feasible connectivity graphs pays high dividends and is the goal of this paper.

For Near-term Intermediate Scale Quantum (NISQ) computers \cite{preskill}, composed of 50-100 physical qubits which function with no or only partial quantum error correction, having optimized schedules makes the difference between being able to run the program and obtain a meaningful result or being left with a noisy and useless output. The ultimate hope of the field is to have error-corrected qubits, each of which consists of many physical qubits, which may be operated for (in-principle) infinitely long times. It is safe to predict that the encoded qubits will also have a limited connectivity graph. Therefore, the research in our paper remains relevant also in the long term by contributing to the very practical task of reducing the execution time of quantum algorithms.

\begin{figure*}[!bht]
    \centering
    \begin{subfigure}[b]{0.45\linewidth}
        \includegraphics[width=\textwidth]{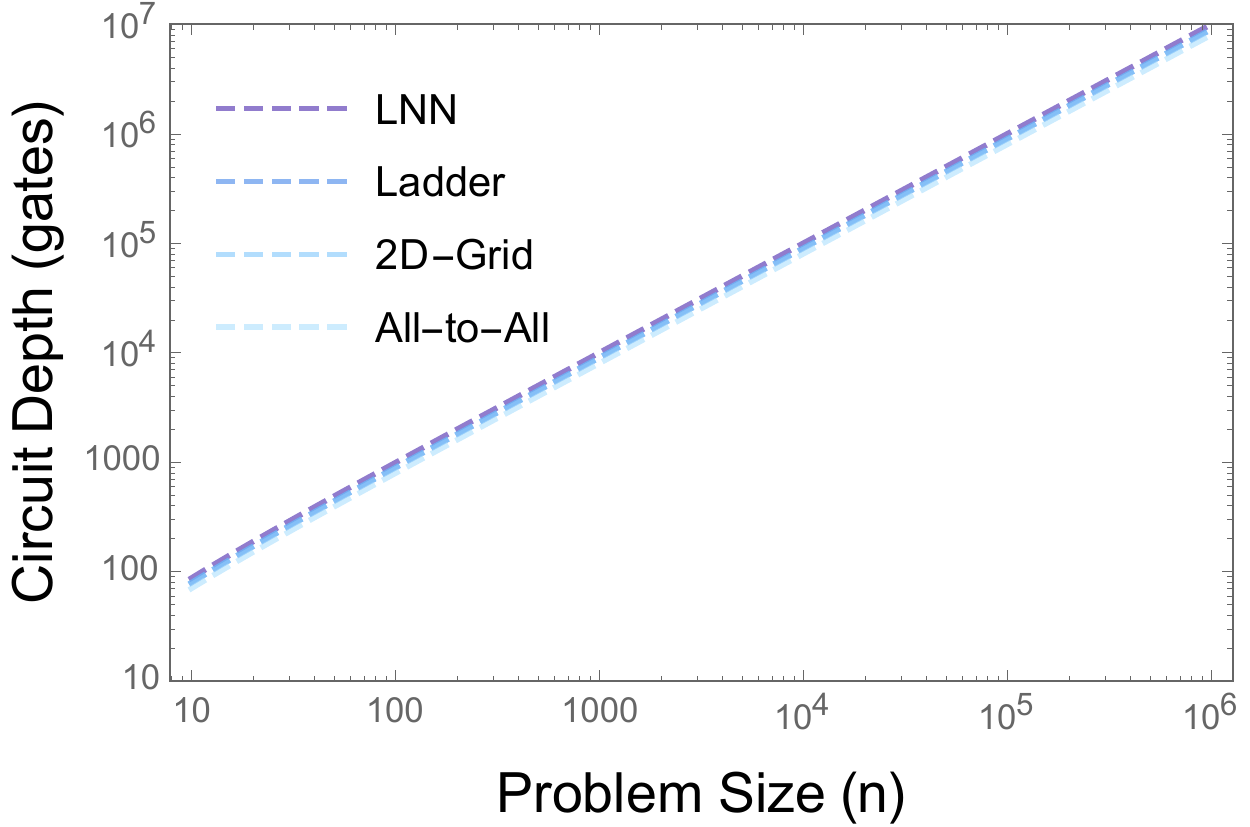}
        \caption{Quantum Fourier Transform}
        \label{fig:qft_results}
    \end{subfigure}
    \begin{subfigure}[b]{0.45\linewidth}
        \includegraphics[width=\textwidth]{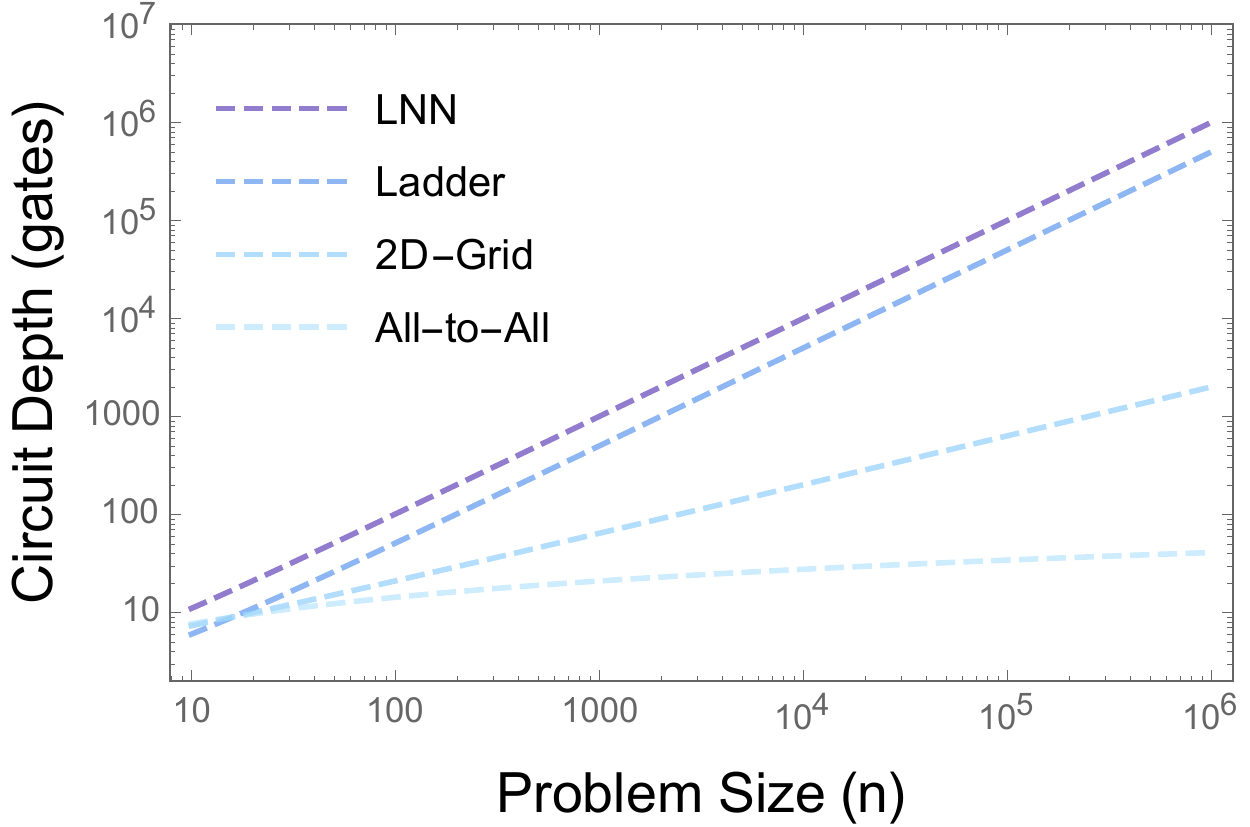}
        \caption{Jordan-Wigner String}
        \label{fig:jw_results}
    \end{subfigure}
    \begin{subfigure}[b]{0.45\linewidth}
        \includegraphics[width=\textwidth]{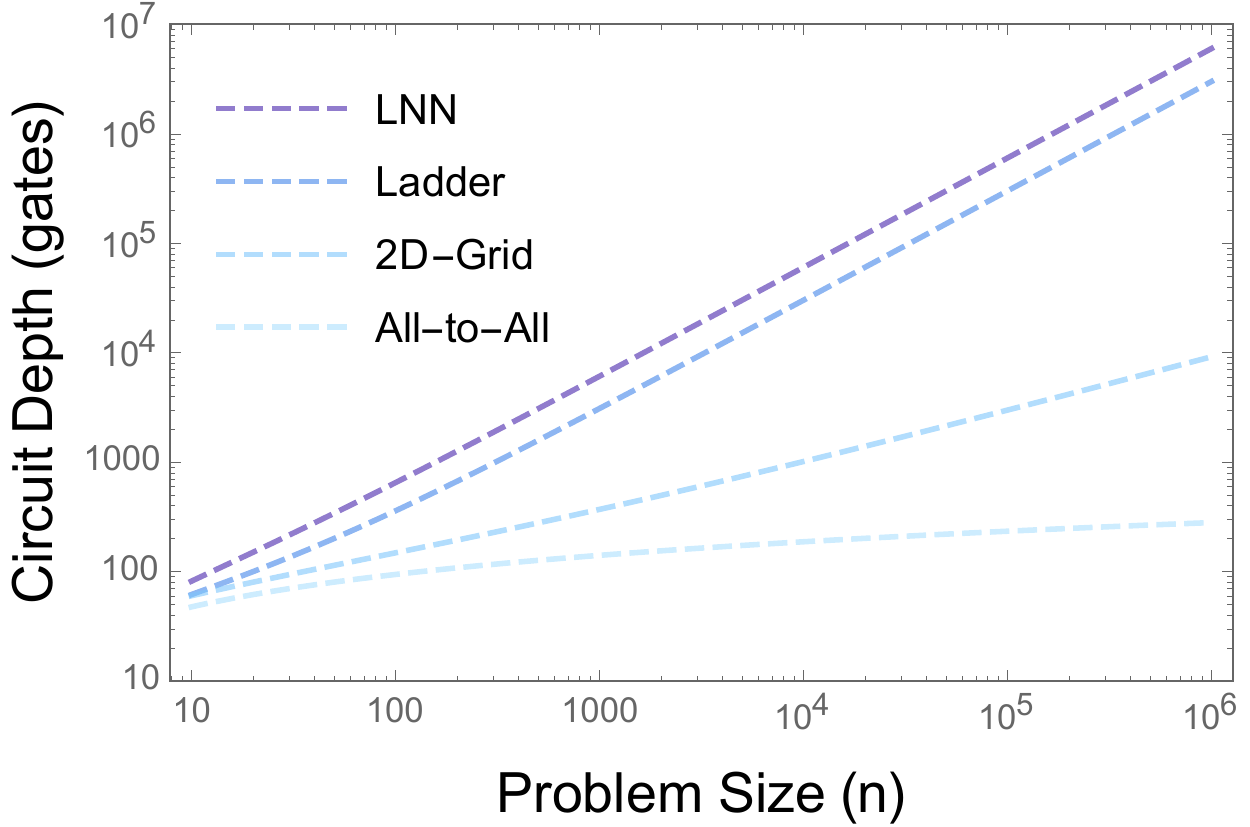}
        \caption{Grover's Diffusion Operator}
        \label{fig:grovers_results}
    \end{subfigure}
    \begin{subfigure}[b]{0.45\linewidth}
        \includegraphics[width=\textwidth]{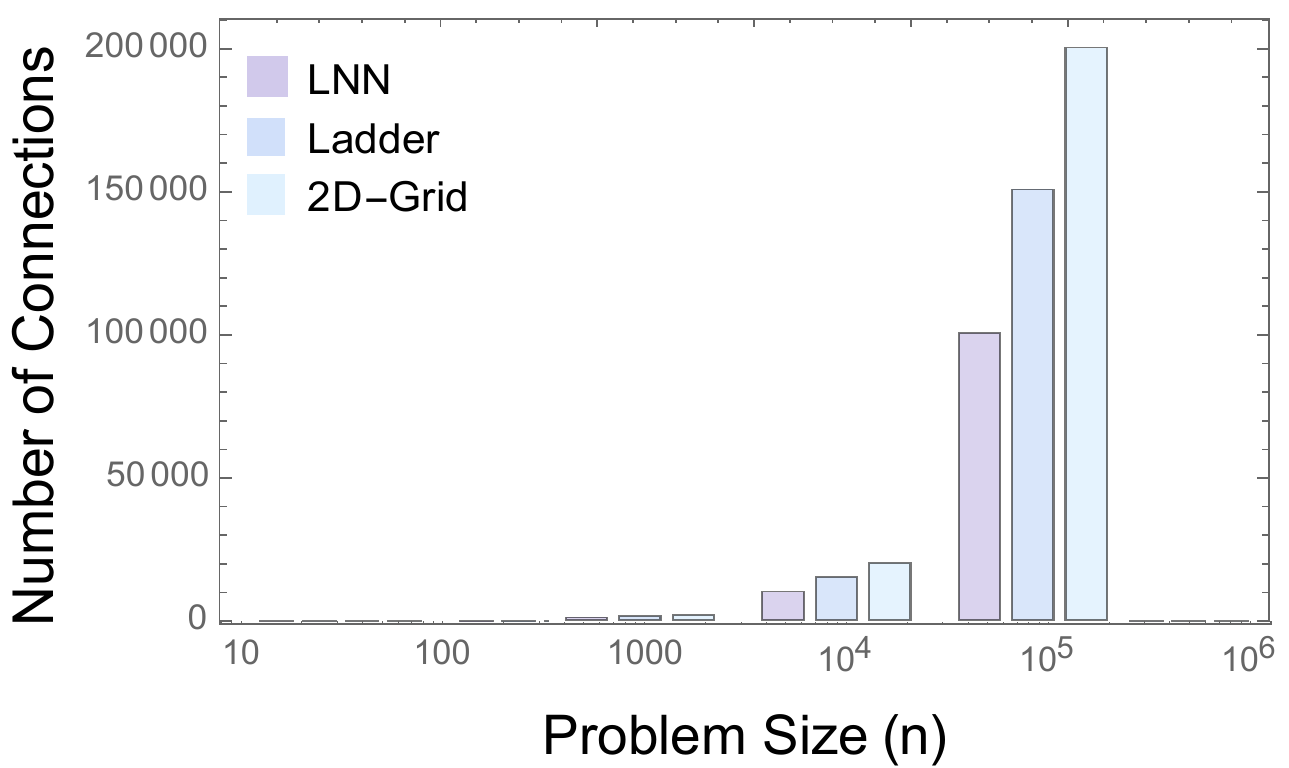}
        \caption{Connection Complexity Scaling}
        \label{fig:grovers_results}
    \end{subfigure}
    \caption{Optimized algorithm performance on various hardware architectures. The precise running times of each algorithm are plotted and overlaid above bar charts denoting the number of physical connections required for each architecture. All-to-all machine performance is shown, but the number of connections is omitted as it is significantly greater than the other machines.}
    \label{fig:results_plots}
\end{figure*}


Other related works in this area have focused on optimizing specific instances of workloads at the application level for specific physical device types \cite{Fowler2004,dorai2005efficient, kivlichan2018quantum}, or for scheduling specific applications onto linearly connected devices \cite{guerreschi2017gate,shafaei2013optimization,saeedi2011synthesis,hirata2009efficient,khan2008cost,wille2014optimal}. A smaller body of work has been developed applying similar scheduling and optimization techniques to two-dimensional qubit plane arrays \cite{Rosenbaum2012,shafaei2014qubit,Brierley2015a}. Additionally, other work has been developed from the perspective of adapting applications to specific physical devices by leveraging techniques from optimal control theory \cite{schulte2007quantum}. 

In this work we analyze the impact of three connectivity graphs, namely a line, ladder, and two-dimensional square grid, on the scaling of important algorithmic workloads like the quantum Fourier transform, Grover diffusion operator, and the parity-based rotations arising from mapping electrons to qubits using the Jordan-Wigner transform. We compare our results with the expected scaling for ideal all-to-all connectivity devices. We conclude that the overhead from routing does not jeopardize quantum speedup even on the linear array, which is the most constrained connectivity graph possible. Note that our emphasis here is on reducing the circuit depth at the cost of adding ancilla qubits. This conclusion is particularly important for near-term devices whose development may prioritize increasing the number of qubits and their quality even before solving the challenges related to higher-dimensional integration. In general, we also observe that common quantum algorithmic primitives exhibit structure that can be exploited to produce much more efficient schedules compared to a random circuit.


A high-level summary of our main results, organized in tables and graphs, is presented in the next section. The rest of the paper covers the required background and detailed results of our study, organized as follows. In section~\ref{sec:background}, we outline the notation that will be adopted throughout the paper. Section~\ref{sec:hardware} introduces the qubit connectivity graphs considered in this study and Section~\ref{sec:algorithms} the algorithms. Section~\ref{sec:results} goes over our results in detail. We conclude with section~\ref{sec:conclusion}. Since our topic is inter-disciplinary, straddling the boundary between physics and computer science, we have provided an introduction to the notation and algorithms we use. Readers familiar with quantum computation can skip directly to section~\ref{sec:results} after section~\ref{sec:summary}.

%% file: summary.tex

\begin{table*}
\centering
\begin{tabular}{@{\extracolsep{4pt}}lcccc@{}}
\hline\hline
Architecture & QFT & Jordan-Wigner String$_{\text{\tiny{\cite{jwapp}}}}$  & Grover's Diffusion Operator\\ \hline
All-to-All & $8n-10_{\text{\tiny{\cite{qftnote}}}}$ & $2\ceil{\log_2{n}}+1$ & $14\log_2{n}+1$ \\ \hline
Linear Nearest Neighbor &$10n-13$ & $n+1 + (n \text{ mod }2)$ & $6n+8\log_2(n)-5$\\
 \hline
Ladder & $9n-11$ &$n/2+1+(n/2 \text{ mod }2)$ & $3n+8\log_2(n)+13$ \\ \hline
2D Grid & $ - $ & $2\sqrt{n}+1+2(\sqrt{n} \text{ mod }2)$ & $9\sqrt{n}+8\log_2(n)+13$ \\
 &  & & \\
\hline\hline
\end{tabular}
\caption{Circuit Depth Results After Gate Decomposition. Here $n$ is the number of qubits involved in each algorithm.
(valid for $n>1).$
}
\label{table:depth}
\end{table*}

Table I shows the main results of our paper. It outlines  the circuit depth under the assumptions of the gate set made in the next section. The main takeaways are:

Quantum Fourier Transform: Though the quantum circuit involves gates between every possible pair of qubits, the circuit depth is still linear in the number of qubits even on a line, which is the same scaling as on an all-to-all connectivity graph. Adding $n/2-1$ connections to form ladder connectivity enables optimization that makes the circuit depth very close to optimal.

Jordan-Wigner String: The Jordan-Wigner string is a building block of fermionic simulation algorithms on quantum computers. On classical computers, these simulations take $\mathcal{O}(\exp{(n)})$ time, where $n$ is the number of qubits or electronic orbitals. On quantum computers, the fermionic simulation algorithms involve a polynomial number of Jordan Wigner strings. The run-time for a single Jordan-Wigner string on qubits with all-to-all connectivity is $\mathcal{O}(\log(n))$. Constraining the connectivity increases the run-time exponentially to become polynomial in $n$. The run-time on a ladder is half that on the line, and the 2D grid is quadratically faster than both of these. 

Grover's Diffusion Operator: In quantum search, the Grover diffusion operator is called after every application of the oracle. For a search space of $N=2^n$, a classical search algorithm calls the oracle $\mathcal{O}(N)$ times, whereas the quantum algorithm calls the oracle $\mathcal{O}(\sqrt{N})$ times. The run-time for the Grover diffusion operator on an all-to-all connectivity qubit device is $\mathcal{O}(\log(n))$ which increases exponentially to be polynomial in $n$ on a constrained connectivity device. Here too, going to a ladder from a line halves the runtime, and the 2D grid is quadratically faster than both of these.

Fig. 1 shows the scaling of these optimized algorithm designs and compares them to the increase in physical complexity of the underlying machines.

%% file: background.tex

In general, a quantum computing algorithm specifies a series of operations on two-level quantum systems called qubits, which are the quantum analogue of classical bits. The state of a single qubit can be represented as a linear combination (superposition) of the two levels as:
\begin{align*}
    \ket{\psi} = \alpha \ket{0} + \beta \ket{1}
\end{align*}
where the linear coefficients $\alpha, \beta \in \mathbb{C}$ are called amplitudes and satisfy $|\alpha|^2 + |\beta|^2 = 1$. During a measurement the state of the qubit collapses into either the $\ket{0}$ or $\ket{1}$ state with probability given by the corresponding amplitude, respectively $|\alpha|^2$ or $|\beta|^2$. The measurement destroys the superposition and return a single bit of information corresponding to which state is obtained at the end of the collapse. 

By extension, a multi-qubit state may be represented as: 
\begin{align*}
    \ket{\psi} = \sum_i \alpha_i \ket{i}
\end{align*}
where $\ket{i}=\ket{i_{n-1}}\otimes\ldots\ket{i_1}\otimes\ket{i_0}$ are the computational basis states of the $n$-qubit quantum system, and $\sum_i|\alpha_i|^2=1$. The symbol $\otimes$ separates the state of the different qubits, as customary in quantum mechanics formalism, and follows from the formal analogy between the composition of quantum systems in the vector/matrix representation and the tensor product. Quantum states can thus be visualized as vectors in a $2^n$ dimensional space. The restriction on the coefficients simply states that they are complex vectors of unit norm.

\subsection{Quantum gates}

Computation is performed by transforming qubit states to different qubit states. Mathematically, any quantum operation corresponds to a unitary transformation that preserves the normalization of the quantum states. When quantum states are visualized like vectors, the transformation corresponds to a unitary matrix $U$ such that $U = U^{\dagger} = (U^*)^T$, where the adjoint ($\dagger)$ indicates the conjugate transpose. 

Despite being a common tool to describe quantum states and transformations, the representation with vectors and matrices will not be used in the following. Instead, we will provide explicit descriptions of how the relevant quantum operations, also called quantum gates, change the quantum states.

\begin{figure*}[t!]
  \centering
  \includegraphics[width=0.75\linewidth]{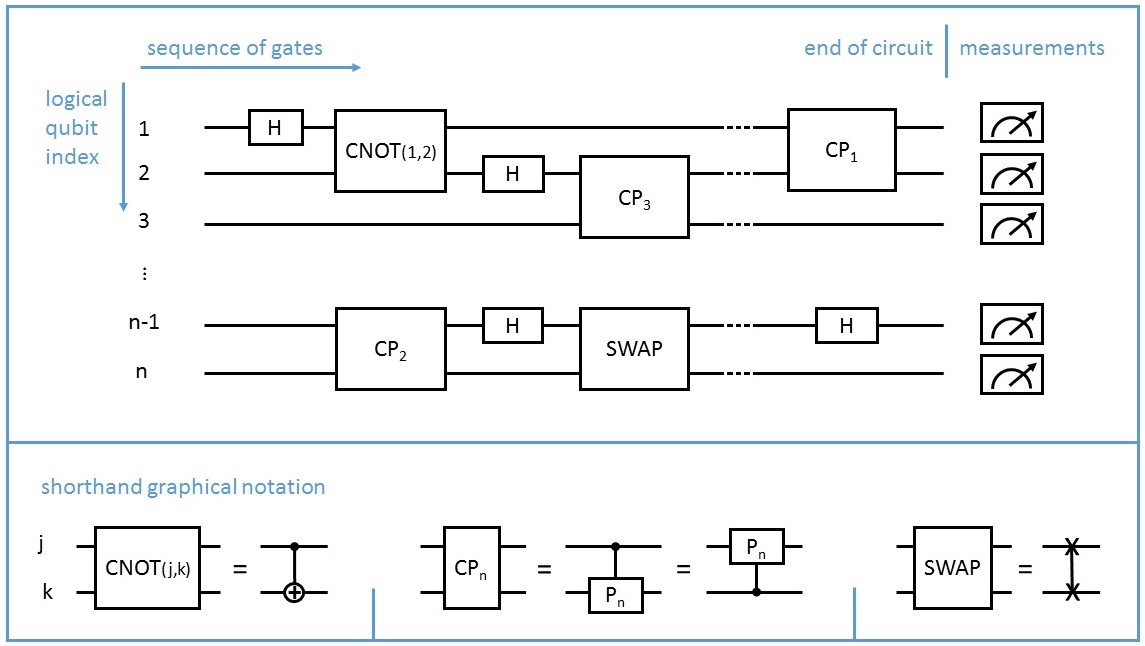}
  \caption{\small{\textbf{Top panel:} example of quantum circuit in which each wire represents a qubit and quantum gates are depicted as boxes. The temporal order of the gates is from left to right and the kind of operation is specified by the name inside the box. At the end of the circuit, all relevant qubits are measured. The graphical representation does not need to include the qubit indices for the gates since the information is already provided by the wires on top of which the boxes are located, with the exception of CNOT for which the qubit order matters. \textbf{Bottom panel:} shorthand notation for the quantum gates introduced in section~\ref{sec:background}.}}
  \label{fig:example-circuit}
\vspace{-0.2cm}
\end{figure*}

The five main operations that we will encounter in the following text are divided into single-qubit operations, like the Hadamard gate H and the Pauli Z gate, and two-qubit operations, namely the controlled-not gate CNOT, the controlled phase CP$_n$ and the SWAP gate. The Hadamard gate H acts on a single qubit as:
\begin{equation}
\label{eq:hadamard}
	\text{H}\,\ket{j} = \frac{\ket{0}+(-1)^j\ket{1}}{\sqrt{2}} \, ,
\end{equation}
on the states $\ket{j}\in\{\ket{0},\ket{1}\}$ and can be extended by linearity to all possible superpositions. The Pauli Z gate corresponds to the transformation:
\begin{equation}
\label{eq:pauliz}
	\text{Z}\,\ket{j} = (-1)^j \ket{j} \,.
\end{equation}

Controlled operations involve two qubits, the first of which determines if the state of the second qubit is modified or left unchanged. Gates of this kind are able to generate entanglement, \emph{i.e.} create quantum correlations between qubits. At least one entangling gate is required to realize a universal quantum computer. The controlled-not gate is described by:
\begin{equation}
\label{eq:cnot}
	\text{CNOT}\,\ket{j}\otimes\ket{k} =
        \ket{j}\otimes\ket{k\oplus j}
\end{equation}
where $j,k\in\{0,1\}$ and the symbol $\oplus$ represents the addition modulo 2. The Toffoli gate (denoted as TOFF) extends this logic, controlling on the state of two qubits and acting on a third. In a similar way, the controlled phase CP$_n$ corresponds to:
\begin{equation}
\label{eq:cphase}
	\text{CP}_n\,\ket{j}\otimes\ket{k} =
	    e^{2 \pi i (k j) /2^n} \ket{j}\otimes\ket{k}
\end{equation}
Finally, the SWAP gate essentially exchanges the state of two qubits, explicitly:
\begin{equation}
\label{eq:swap}
	\text{SWAP}\,\ket{j}\otimes\ket{k} = \ket{k}\otimes\ket{j} \, .
\end{equation}

We specify which qubits are involved in the gate by providing the qubit indices as functional arguments to the gate name, for example H$(n)$ acts on qubit $n$ and CNOT$(n,m)$ modifies target qubit $m$ depending on the state of the control qubit $n$. Notice that CP$_n$ and SWAP act on two qubits in a symmetric way, thus the order of the qubit indices is not important for this operations, \emph{e.g.} SWAP$(n,m$)=SWAP$(m,n)$.

In order to standardize our analysis across applications, we will be using a decomposition of each of the controlled phase gates into 3 single-qubit phase gates and 2 CNOT gates. The set of available gates is provided at the beginning of section~\ref{sec:results}. The controlled-phase gate decomposition is shown at the bottom of FIG.~\ref{fig:qftcirc} and derived in Appendix \ref{sec:gate-decomposition-cp}, while a Toffoli gate decomposition is described in Appendix \ref{sec:gate-decomposition-toffoli}.

\subsection{Quantum Algorithms}

A quantum algorithm is fundamentally a unitary transformation being performed on a set of $n$ qubits. In order to physically implement such unitary transformations, it is desirable to decompose algorithms into a sequence of operations so that no individual unitary transformation is operating on more than two (in some cases,  three) qubits at a single time. It is possible to find such decompositions systematically, as shown in \cite{mikeike}. 

The sequence of few-qubit gates can be depicted by quantum circuits, which are temporally ordered sequences of quantum gates acting on one or more qubits. Figure~\ref{fig:example-circuit} shows an example. Here, each horizontal line (or wire) depicts a single qubit and each box indicates a quantum gate. The most used gates are usually represented with dedicated graphical symbols as provided in the bottom panel of the same figure.

Running such a quantum circuit on an actual quantum information processor requires that each gate is physically realizable. This imposes a \textit{connectivity} constraint for two-qubit gates. More precisely, given a two-qubit unitary transformation $U$ to be performed as part of a quantum algorithm, we require at the time of execution that the two qubits involved physically interact in some way.

Accommodating this constraint can be achieved in a few ways, a common choice being the utilization of SWAP operations. As described in equation~\eqref{eq:swap}, this two-qubit transformation exchanges the information content of two qubits, effectively changing the logical-to-physical mapping that indicates what logical qubit (from the algorithm) is associated with what physical qubit (in the hardware). By inserting chains of SWAP gates, logical qubit states can be routed throughout the physical system, allowing the locality constraint to be satisfied for all the two-qubit gates of the algorithm.

\begin{figure*}[t!]
  \centering
  \includegraphics[width=0.7\linewidth]{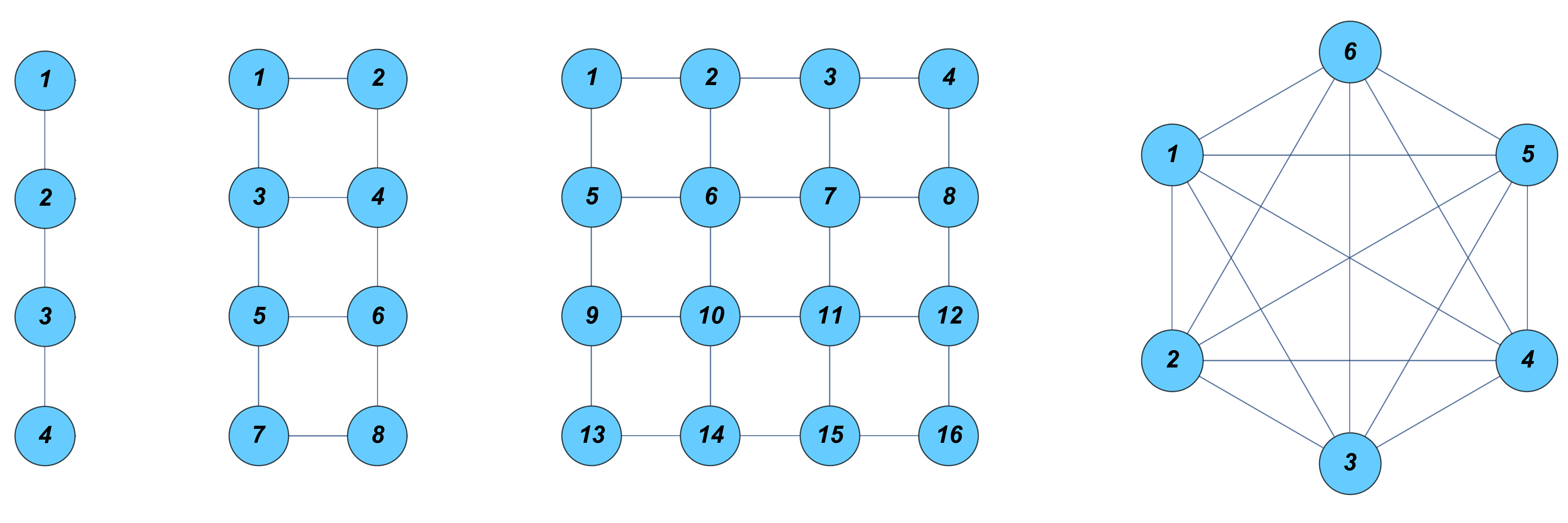}
  \caption{Connectivity graphs of  (from left to right) linear, ladder, grid, and all-to-all connected devices. Physical qubits are represented by the nodes and the edges corresponds to the possible locations of two-qubit gates. The labels indicate the physical qubit indices which will be used in our algorithm presentations.} 
  \label{fig:hardware}
\end{figure*}

A given physical architecture will have a limited gate set it can implement. Compiling to this gate set gives the total number of gates. A further set of constraints will arise which dictate which gates can be executed in parallel. This will define the circuit depth which can be thought of as the time to execute the circuit if each gate takes time 1 in arbitrary units. In this paper, we assume there are no control constraints and try to minimize the circuit depth.

%% file: hardware.tex
Limited connectivity graphs can impose a large overhead if several SWAP operations are required in order to complete the execution of every gate. To quantify the impact of the connectivity, we consider four typical hardware architectures that seem feasible for physical realization and describe them as connectivity graphs (here each node represents a physical qubit and each edge indicates the availability of two-qubit gates between the connected qubits). For a total of $n$ qubits, and from the most constrained to the least constrained connectivity, one has:
\begin{description}
    \item[linear] Linear nearest neighbor graph with open boundaries and $n-1$ edges. In solid-state implementations, qubit control lines here could be in the same plane making this easy to manufacture. Realized with trapped ions \cite{haffner2005scalable}, superconducting qubits \cite{Kelly2015,barends2014superconducting} and quantum dots \cite{zajac2016scalable}.
    \item[ladder] Grid of dimensions $n/2 \times 2$ with $3n/2-2$ edges. One can visualize it as comprised of two columns of $n/2$ qubits each \cite{ibmsite}.
    \item[square grid] Square grid of dimensions $\sqrt{n}\times\sqrt{n}$ with $2(n-\sqrt{n})$ edges. An extension of this design is naturally suitable for topological error correction, e.g. by encoding information via the surface code \cite{googlesite,li2017crossbar}. It likely requires the development of out-of-plane control lines to address the qubits in bulk.
    \item[all-to-all] Fully connected graph with $n(n-1)$ edges. No need for routing. It has been implemented in small trapped ion systems \cite{ti_monroe}, but likely infeasible for large number of qubits. Often used in the abstract description of algorithms.
\end{description}

If the number of qubits is such that a ladder or square grid is not fully filled, one can consider the smallest integer larger than $n$ that would complete the structure in the above estimates. In addition, all of these designs are assumed to provide complete and independent qubit control, meaning that one can perform gates in parallel if and only if they act on different qubits.

As indicated in the description, these architectural design choices are also intended to sweep the physical fabrication complexity spectrum, with the most constrained connectivity graphs being the easiest to fabricate. Graphically, the architectures are shown in Figure~\ref{fig:hardware}. In this work we analyze the performance of each of these architectures with respect to three important quantum applications and subroutines, described in the next section.


%% file: algorithms.tex
\begin{figure*}[t!]
  \centering
  \includegraphics[width=0.8\linewidth]{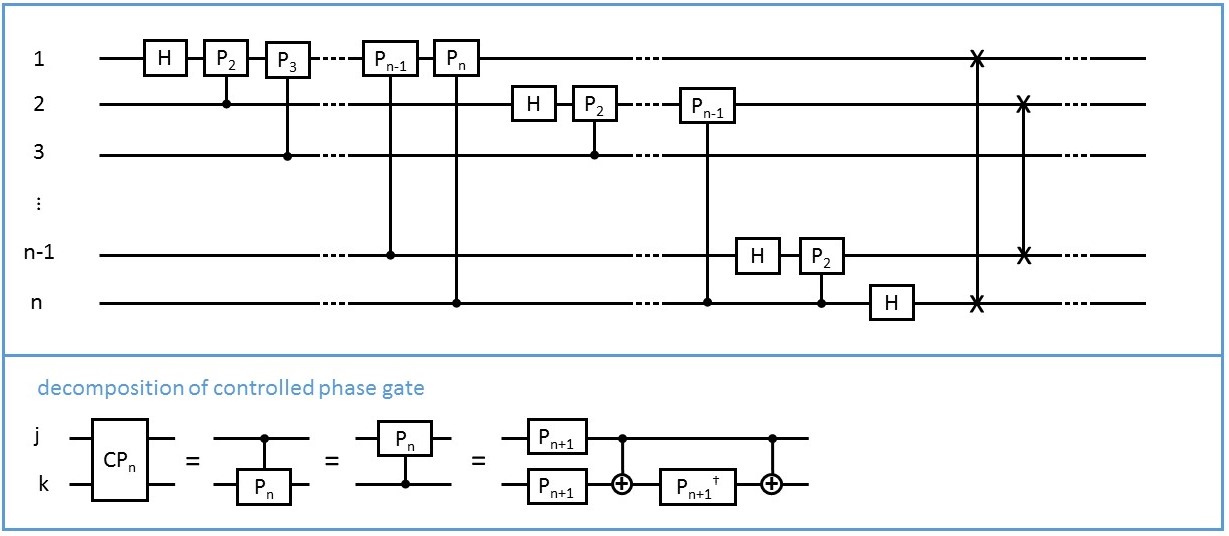}
  \caption{\small{Circuit implementation of the Quantum Fourier Transform from reference \cite{mikeike}. The SWAP gates at the end of the circuit are used to invert the qubit order and restore the logical one from Eq.~\eqref{eq:QFT}. The explicit decomposition of the controlled phase gate into single-qubit phase gates and CNOT gates is included in the bottom panel.}}
  \label{fig:qftcirc}
\end{figure*}

Two goals are balanced in the selection of the algorithms to study: representation of a large portion of all quantum algorithms, and algorithmic complexity class separation. Specifically, we select quantum subroutines that are present as a significant component of many other quantum algorithms so as to cover a significant portion of the benchmarks. Additionally we seek to ensure that algorithms with both \textit{exponential} and \textit{polynomial} speedup, with respect to their counterpart classical algorithms, were represented. This choice allows our results to speak to the impact that particular hardware architectural constraints impose upon quantum speedups at different scales.


\subsection{Quantum Fourier Transform}

The Quantum Fourier Transform (QFT) \cite{mikeike} is the quantum analog of the classical discrete Fourier transform, in which an input vector of complex numbers $\bx = (x_0, x_1, ... , x_{N-1})$ is transformed to a vector of complex numbers $\by = (y_0, y_1, ... , y_{N-1})$ according to:
	\begin{equation}
	   \label{eq:FFT}
		y_k = \frac{1}{\sqrt{N}}\sum_{j=0}^{N-1} e^{2\pi ijk/N}\,x_j	
	\end{equation} 
The Quantum Fourier Transform takes an input state from the computational basis, namely $\{ \ket{0}, \dots , \ket{N-1}\}$, and transforms it according to:
	\begin{equation}
	   \label{eq:QFT}
		\ket{k} = \frac{1}{\sqrt{N}}\sum_{j=0}^{N-1} e^{2\pi i j k/N}\,\ket{j}
	\end{equation}
For $N=2^n$, the circuit representation of the QFT requires $n$ qubits and is provided in Figure~\ref{fig:qftcirc} following the usual description \cite{mikeike}. Note that the most significant bit corresponds to the upper most qubit in Figure~\ref{fig:qftcirc}: $(n-1)$ controlled phase gates act on logical qubit 1, $(n-2)$ on logical qubit $2$ and, in general, $(n-k)$ on logical qubit $k$, combining to a total $\mathcal{O}(n^2)$ gates to perform the transformation on $2^n$ numbers. While the number of gates is $\mathcal{O}(n^2)$, the depth of the circuit on an all-to-all connectivity graph can be shown to be $\mathcal{O}(n)$ 
which sets the theoretical lower bound of any schedule of the operations of the QFT.



As a quantum subroutine, the QFT is an important component of many quantum algorithms including Shor's prime factorization and discrete logarithm algorithms \cite{Shor1999}, the estimation of eigenvalues and eigenvectors of matrices \cite{Abrams1999}, and others. In particular, the QFT can be seen as a general routine needed to solve problems that can be reduced to the \textit{hidden subgroup problem}, of which the listed examples above are special cases.

The speedup from the QFT comes from the comparison with analogous classical algorithms for performing the Fast Fourier Transform of $N = 2^n$ numbers. The best known algorithm to perform this computation requires roughly $\mathcal{O}(N \log N) = \mathcal{O}(n 2^n)$ operations, while invoking the quantum Fourier transform on the same input only requires approximately $\mathcal{O}( \log^2 N) = \mathcal{O}(n^2)$ operations, resulting in an \textit{exponential} speedup.

\subsection{Jordan-Wigner String:\\ Rotations based on parity operators}
The study of materials and molecules is intimately related to the solution of the associated electronic problem: simulating fermionic systems is exponentially expensive on classical computers, but can be efficiently performed with quantum computers \cite{Ortiz2001,whit}. In fact, quantum chemistry is often regarded as the \emph{killer application} of near term quantum devices \cite{ceen}.

The Jordan-Wigner transform is a method by which a system of fermions, most notably electrons, can be mapped onto a system of qubits. In the mapping, the number of electronic ``orbitals'' is fixed to be a constant. Due to the Pauli exclusion principle, at most one electron can occupy an orbital. One can assign a qubit to each orbital and associate state $\ket{0}$ to the absence of the electron and state $\ket{1}$ to the occupied orbital. However, fermionic states must satisfy anti-symmetry constraints known as the Fermi-Dirac statistics. From a practical perspective, if one wants to create an electron in a certain orbital, a minus sign may be required depending on the occupancy of the other orbitals: to add an electron in the $k$-th orbital, the minus sign is necessary if an odd number of electrons are present in the orbitals with index $j<k$.

The most expensive step in fermionic simulations arises from the parity operation needed to implement Fermi-Dirac statistics:
\begin{equation}
\label{eq:parity}
	\text{PAR}\,\ket{j}\otimes\dots\otimes\ket{k} =
	    (-1)^{j\oplus\dots\oplus k} \ket{j}\otimes\dots\otimes\ket{k}
\end{equation}
that introduces a minus sign depending on whether the number of involved qubits that are in state $\ket{1}$ is odd or even. 

\begin{figure}[t!]
\begin{center}
 \includegraphics[width=0.95\linewidth]{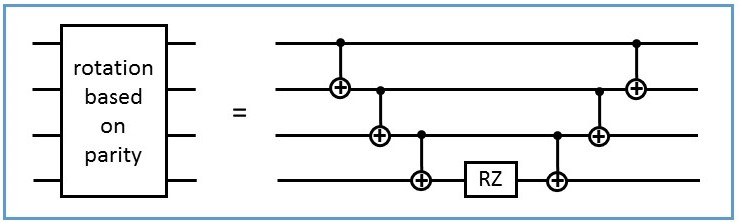}
\caption{\small{Quantum circuit implementing a typical operator used in quantum simulations of fermionic systems via the Jordan-Wigner transformation. Specifically, the circuit corresponds to the four-qubit operation $\exp{(-i \theta/2\,\text{PAR})}$, with PAR = $Z_1 \tens Z_2 \tens Z_3 \tens Z_4$.}}
\label{fig:jw}
\end{center}
\end{figure}

For a single qubit PAR=Z, while in full generality it may involve all qubits.
We call ``rotation based on parity'', or a ``Jordan-Wigner string", the operation described by:
\begin{align}
\label{eq:parityrotation}
	\exp\left(-i \tfrac{\theta}{2} \text{PAR}\right)\,\ket{j}\dots\ket{k} =
	\hspace{3cm}\nonumber\\ =
        \begin{cases}
            e^{-i \pi \tfrac{\theta}{2}} \ket{j}\dots\ket{k}
                & \mbox{if } j\oplus\dots\oplus k=0\\
            e^{+i \pi \tfrac{\theta}{2}} \ket{j}\dots\ket{k}
                & \mbox{if } j\oplus\dots\oplus k=1
        \end{cases} \, ,	
\end{align}
where $\theta$ is an arbitrary angle. In the following, rotations based on parity are considered the appropriate proxy operation to implement fermionic simulations%
\footnote{The reader familiar with quantum simulation will realize that each factor of the Trotter-Suzuki decomposition of the propagator can be written as a rotation based on parity when suitable single-qubit operators are added to change the local basis.}.
The decomposition of the rotation into one- and two-qubit gates typically takes the form of a ``CNOT staircase'' illustrated in Figure~\ref{fig:jw}, where RZ$(\theta)=e^{-i\frac{\theta}{2} Z}$.

Observe that the local basis of each qubit can be changed before and after the parity-based rotation, in this way the circuit implements rotations based on arbitrary products of Pauli operators. The overhead due to a change in basis in circuit depth is minimal and $\mathcal{O}(1)$ irrespective of the number of qubits $n$.

Finally we observe that, while the CNOT staircase is intuitively simple, without limited connectivity one can dramatically reduce the circuit depth from $\mathcal{O}(n)$ to $\mathcal{O}(\log_2 n)$ by parallelizing the CNOTs according to a tree structure.

\subsection{Grover Diffusion Operator}
The previous two quantum routines are part of quantum algorithms providing exponential speedup over their classical equivalents. To represent quantum algorithms with \textit{polynomial} speedup relative to their classical counterpart, we consider the task of searching for a target element inside an unstructured database. Grover's quantum search algorithm \cite{Grover1997} can be described with the following: Given a search space of size $N$ and no prior knowledge of the structure of the space, find the unique target element. Classically, $\mathcal{O}(N)$ queries are required to find the target element on average, while only $\mathcal{O}(\sqrt{N})$ queries are required with quantum search. 

The algorithm begins with initialization of the input into a uniform superposition state:
\begin{equation}
	\ket{\psi} = \frac{1}{\sqrt{N}}\sum_{j=0}^{N-1}\ket{j} \; ,
\end{equation}
where $\ket{j}$ are computational basis states of the $N$ dimensional space, and $j$ can be seen in binary form as the index of a $\log_2{N}$-qubit state.

The algorithm is characterized by the \textit{Grover iteration} where, in addition to the oracle operation corresponding to a classical query of the database, the only non-trivial operation is the so-called \textit{Grover diffusion operator}. In practice, it is a conditional phase shift that requires the knowledge of the state of all qubits and its implementation captures most of the complexity of the algorithm apart from the oracle implementation. In abstract terms, the Grover diffusion operator applies a minus sign to a specific quantum state and takes the form of:
\begin{equation}
	U = \bone - 2\ket{N-1}\bra{N-1}
\end{equation}
where $\ket{N-1}=\ket{1}\otimes\ket{1}\dots\otimes\ket{1}$ in qubit language. In part of the literature, the role of the only state that acquires the minus sign is played by $\ket{0}=\ket{0}\otimes\ket{0}\dots\otimes\ket{0}$ instead, the difference being the introduction of an extra layer of $X$ gates and an $O(1)$ increase of circuit depth.

For practical implementations, such a multi-qubit operation needs to be decomposed in terms of gates involving at most two qubits at a time.  More precisely, the key operation is the unitary version of the classical circuit that computes the logical AND of $n=\log_2{N}$ bits (this is valid for the correspondence $\ket{0}\rightarrow$False and $\ket{1}\rightarrow$True). These circuits are reversibly comprised of a sequence of Toffoli gates (see next paragraph), and are characterized by gate complexity of order $\mathcal{O}(\log N)$.

To count the number of gates, we will make use of a decomposition of the Toffoli gate into 3 CNOT operations and 4 single qubit rotations, following the construction of Margolus in section VI.B of \cite{Barenco}, and reported in Appendix \ref{sec:gate-decomposition-toffoli}. This construction is congruent to the canonical Toffoli gate modulo a phase shift ($\ket{101} \rightarrow -\ket{101}$). The accumulated phase is canceled before proceeding on to the next component of the algorithm. This decomposition introduces two qubit gates between the control qubits and the target qubit, which induces additional SWAP gates if the controls are not both physically adjacent to the target qubit.  


As these circuits are repeatedly required for execution of any instantiation of quantum search or function inversion, they represent a wide range of quantum algorithms that show a \textit{polynomial} speedup over the classical counterparts \cite{Brassard2002}.

%% file: results.tex

In this section, 
we map and schedule the benchmarks discussed above on each of the qubit connectivity graphs under consideration. These algorithms are presented in high level overviews, followed by pseudocode descriptions, followed lastly by analysis.


We use the convention that any operation contained in a block labeled by \textbf{parallel} will be executed in parallel with all of the other gates contained in the \textit{unrolled} version of the block. Blocks may include conditional or loop statements, like ``while'' and ``for'' instructions. Unrolling the blocks amounts to resolving all indices of any operation contained in the block, and inlining the resulting gates.  

To standardize the analysis across these benchmarks, we describe all circuits in terms of a native gate set composed by the Hadamard gate, single-qubit rotations (also equivalent to single qubit phase gates apart from an unobservable global phase), and the CNOT gate. All of the algorithmic benchmarks are decomposed into this basis, and gate counts and circuit depths are reported in terms of these operations. Observe that the SWAP operation is decomposed into a sequence of three CNOT gates, while the controlled phase and Toffoli gate are decomposed as indicated in the descriptions of the benchmarks in section \ref{sec:algorithms}. These decompositions are utilized in resource overhead calculations, but are omitted in pseudocode algorithm descriptions for clarity.

\subsection{Linear Array Results}

\subsubsection{Quantum Fourier Transform}


We first show that the Quantum Fourier Transform on a linear array requires $10n-13$ operation steps. A similar approach was presented in \cite{Fowler2004} as part of the circuit implementing Shor's algorithm, but with a gate set containing arbitrary 2-qubit gates.

\begin{figure*}
\begin{center}
 \includegraphics[width=\textwidth]{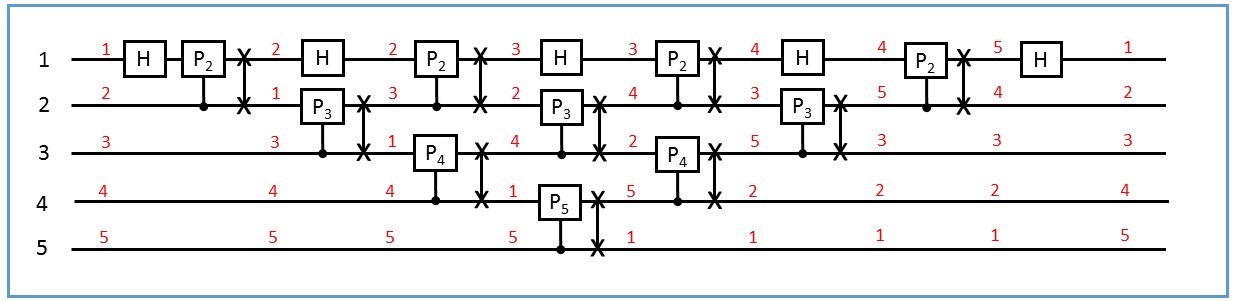}
\caption{\small{The Quantum Fourier Transform on a linear array, as described by Algorithm \ref{alg:qftlnn}. The physical index of the qubits is noted in black at the left end of the quantum circuit, while the red labels correspond to the logical indices which vary during the circuit due to the extra SWAP operations. Time runs from left to right.}}
\label{fig:schedule-qft-lnn}
\end{center}
\end{figure*}

The initial placement of the qubits is the trivial one, with each logical qubit $x_i$ mapped to the physical qubit $q_i$. We order the qubits from top to bottom as in Fig. \ref{fig:hardware}. The implementation proceeds by performing the first Hadamard on the top qubit (index 1), followed by a controlled-phase gate between this qubit with its neighbor. This is followed by a SWAP between the top two qubits. The sequence is repeated on the top qubit which now has logical index 2. Meanwhile logical qubit 1 is available for a controlled-phase gate with logical qubit 3. With each repetition of this pattern, the number of SWAP and CP gates performed in parallel is increased by one. The scheme is presented in Algorithm \ref{alg:qftlnn}, and visualized in Fig. \ref{fig:schedule-qft-lnn}.

\begin{algorithm}[H]
	\caption{Linear Quantum Fourier Transform}
	\label{alg:qftlnn}
	\begin{algorithmic}[1]
	    \item [logical qubit $x_i$ mapped to physical qubit $q_i$]
		\State $\text{H } (q_1)$
		\For{$i$ in $\{2,3,4,\dots,n-1,n,n-1,\dots,2\}$}
			\State $j \gets i$
			\While{$j \geq 2$} \textbf{Parallel}
				\State $\text{CP}_j \ (q_j, q_{j-1})$ 
				\State $j \gets j - 2$
				\If{$j == 1$}
					\State $\text{H } (q_1)$
				\EndIf
			\EndWhile
			\State $j \gets i$
			\While{$j \geq 2$} \textbf{Parallel}
				\State $\text{SWAP } (q_j, q_{j-1})$ 
				\State $j \gets j - 2$
			\EndWhile
		\EndFor
		\State $\text{H } (q_1)$
	\end{algorithmic}
\end{algorithm}

	\par \textbf{Performance Analysis}: In the structure of the algorithm, there is a single outer loop iterating $2n-3$ times. Each iteration of this loop is comprised of two separate inner loops. Each of these loops is completely parallelizable, such that the circuit depth of any iteration of the outer loop is equal to the summation of the depth required to execute a single gate from both of the inner loops, which after gate decomposition requires a total of $7$ time steps. Notice however that a single $CP_k$ gate followed by a SWAP gate on the same qubit operands offers a circuit optimization in which two inner CNOT operations of the decomposed version of this gate sequence cancel. Because of this, the depth of a single iteration of the outer loop is reduced by 2. Added to this are the leading and final Hadamard operations, resulting in a final running time of $10n-13$. 

	

\subsubsection{Jordan-Wigner String}
    The Jordan-Wigner transform is analyzed using the proxy benchmark of rotations based on parity operators. For a quantum machine with linear nearest neighbor qubit connectivity, any parity-based rotation can be implemented with a circuit of depth at most $(n+1)$ or $(n+2)$, respectively for $n$ even or odd. This includes rotations based on any product of Pauli operators acting on any subset of qubits within the machine.
    
    \begin{algorithm}[H]
    	\caption{Linear Jordan-Wigner String: $e^{-i\frac{\theta}{2}\sigma_z^{\tens_n}}$}
    	\label{alg:jw1sched}
    	\begin{algorithmic}[1]
    		\item [arbitrary map between logical and physical qubits]
    		\State $m \gets \lfloor{n/2}\rfloor$
    		\State $k \gets n\pmod{2}$
    		\State $i \gets 1$
    		\While{$i < m $} \textbf{Parallel}
    		\State \text{CNOT } $(q_i, q_{i+1})$
    		\State \text{CNOT } $(q_{n-i+1}, q_{n-i})$
    		\State $i \gets i + 1$
    		\EndWhile
    		\State \text{CNOT } $(q_{m+1+k}, q_{m+k})$
    		\If{$k==1$}
    		\State \text{CNOT } $(q_{m}, q_{m+1})$
    		\EndIf
    		\State \text{Rz } ($q_{m+k}, \theta$)
    		\If{$k==1$}
    		\State \text{CNOT } $(q_{m}, q_{m+1})$
    		\EndIf
    		\State \text{CNOT } $(q_{m+1+k}, q_{m+k})$
    		\State $i \gets i - 1$
    		\While{$i > 0$} \textbf{Parallel}
    		\State \text{CNOT } ($q_i, q_{i+1})$
    		\State \text{CNOT } ($q_{n-i+1}, q_{n-i})$
    		\State $i \gets i - 1$
    		\EndWhile
    	\end{algorithmic}
    \end{algorithm}
    
\begin{figure*}
\begin{center}
 \includegraphics[width=\textwidth]{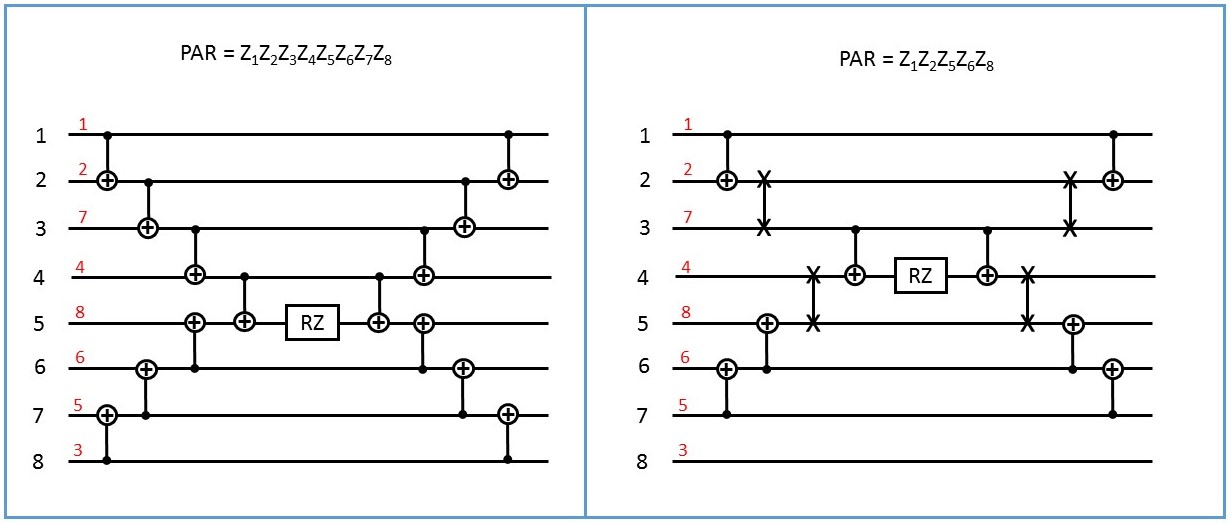}
\caption{\small{Schedule for the parity-based rotations with 8 qubits. The physical index of the qubits is noted in black, while the red label corresponds to the logical index. Initially they coincide, but the application of SWAP gates make them differ along the circuit. Left panel: circuit that calculates global parity. Right panel: parity only involves logical qubits $\{x_1,x_2,x_5,x_6,x_8\}$.}}
\label{fig:schedule-1d-jw}
\end{center}
\end{figure*}

	We consider two cases for the PAR operator, depending on whether it involves all or a subset of the qubits. In the first case the optimal approach corresponds to the CNOT staircase presented in section~\ref{sec:background}.B, but starting from both the end points of the chain and performing the rotation on the physically central qubit. Pseudo-code for this construction is provided in Algorithm~\ref{alg:jw1sched}, and a visualization is included in Fig.~\ref{fig:schedule-1d-jw}. Notice that it works irrespective of how the logical qubits are mapped to the physical register. 
	
	For future reference we refer to the first half of Algorithm \ref{alg:jw1sched} by PARITY, and its inverse operation by PARITY$^{\dagger}$. This subroutine stores the parity of the involved qubits in the central qubit, and will be used as a component of later algorithmic constructions.

More interesting is the situation where PAR involves a randomized subset of the qubits, possibly spread over the full machine and including physically disconnected qubits. Consider the physical-to-logical index map $f:\mathbb{N}\rightarrow\mathbb{N}$ such that $f(i)=j$ indicates that physical qubit $q_i$ is associated to logical qubit $x_j$. In addition, we introduce the function $p:\mathbb{N}\rightarrow\{0,1\}$ that for each index $j$ returns $p(j)=1$ if logical qubit $x_j$ is part of PAR or $p(j)=0$ otherwise. 

\begin{algorithm}[H]
    \caption{PARITY Subroutine}
    \label{alg:PARITY}
    \begin{algorithmic}[1]
    \State $m \gets \lfloor{n/2}\rfloor$
	\State $k \gets n\pmod{2}$
	\State $i \gets 1$
    \While{$i < m $} \textbf{Parallel}
			\State \text{CNOT } $(q_i, q_{i+1})$
    		\State \text{CNOT } $(q_{n-i+1}, q_{n-i})$
			\State $i \gets i + 1$
		\EndWhile
        \State \text{CNOT } $(q_{m+1+k}, q_{m+k})$
        \If{$k==1$}
            \State \text{CNOT } $(q_{m}, q_{m+1})$
        \EndIf
    \end{algorithmic}
\end{algorithm}

With these definitions, it is straightforward to modify the pseudocode to the new situation. In practice, substitute each CNOT$(q_i,q_j)$ operation for which $p(f(j))=0$ with SWAP$(q_i,q_j)$. The circuit can be optimized by eliminating initial SWAPs at the edge of the chain between physical qubits not involved in PAR. Finally, since the duration of CNOT and SWAP may differ (for example, a single SWAP is usually decomposed in terms of 3 consecutive CNOT gates), it is better to execute the gates from the end points in parallel even if in an asynchronous manner.



\textbf{Performance Analysis}: Walking through the algorithm, the general case is executed with complexity $\mathcal{O}(n)$. The worst case performance is dependent upon the relative gate durations of the CNOT and the SWAP operations: if CNOTs require more time overhead than SWAPs, the worst case is achieved when PAR involves all qubits. If instead SWAPs take longer to execute than CNOTs, the worst case occurs when PAR involves only the two end-points of the linear machine respectively. Formally, the latter situation corresponds to PAR=$Z_i\tens Z_j$ with $\{i,j\}=\{f(1),f(n)\}$. 
The exact cost depends on the distance between the two farthest ``active'' qubits in the chain and the ratio between the required CNOT and SWAP gates. In cases with required SWAP operations, the running time can be parameterized directly by inlining the gate decomposition and counting these operations as $n+1+(n \text{ mod } 2)$, or $ N_{\text{CNOT}} + 3N_{\text{SWAP}} +1+(n \text{ mod } 2)$, where $N_{\text{CNOT}}$ and $N_{\text{SWAP}}$ denote the total number of logical CNOT and SWAP operations in the algorithm, respectively. In the results table~\ref{table:depth}, the depth is reported for an instantiation of the algorithm with global parity.

\subsubsection{Grover Diffusion Operator}
\label{alg:groverline}

As mentioned in section \ref{sec:algorithms}, the main subroutine of Grover's quantum search algorithm computes the logical AND of $n$ bits using a sequence of $\mathcal{O}(\log n)$ Toffoli gates, each of which has a small constant expansion into one and two qubit gates. To schedule this circuit, consider a \textit{binary tree} containing a total of $(2n-1)$ nodes: the $n$ leaf nodes represent the data qubits, while all other nodes correspond to ancilla qubits initialized in state $\ket{0}$. The desired Grover diffusion operator can be realized by operating from the leaf nodes up through the tree, by performing a Toffoli gate targeting the root of any subtree and controlled by the two children nodes of the subtree. Each intermediate node contains the result of the logical AND between all leaves of the subtree that has it as the root, and therefore the unique root of the full binary tree contains the logical AND of all the data qubits.


\begin{figure}
\begin{center}
 \includegraphics[width=0.95\linewidth]{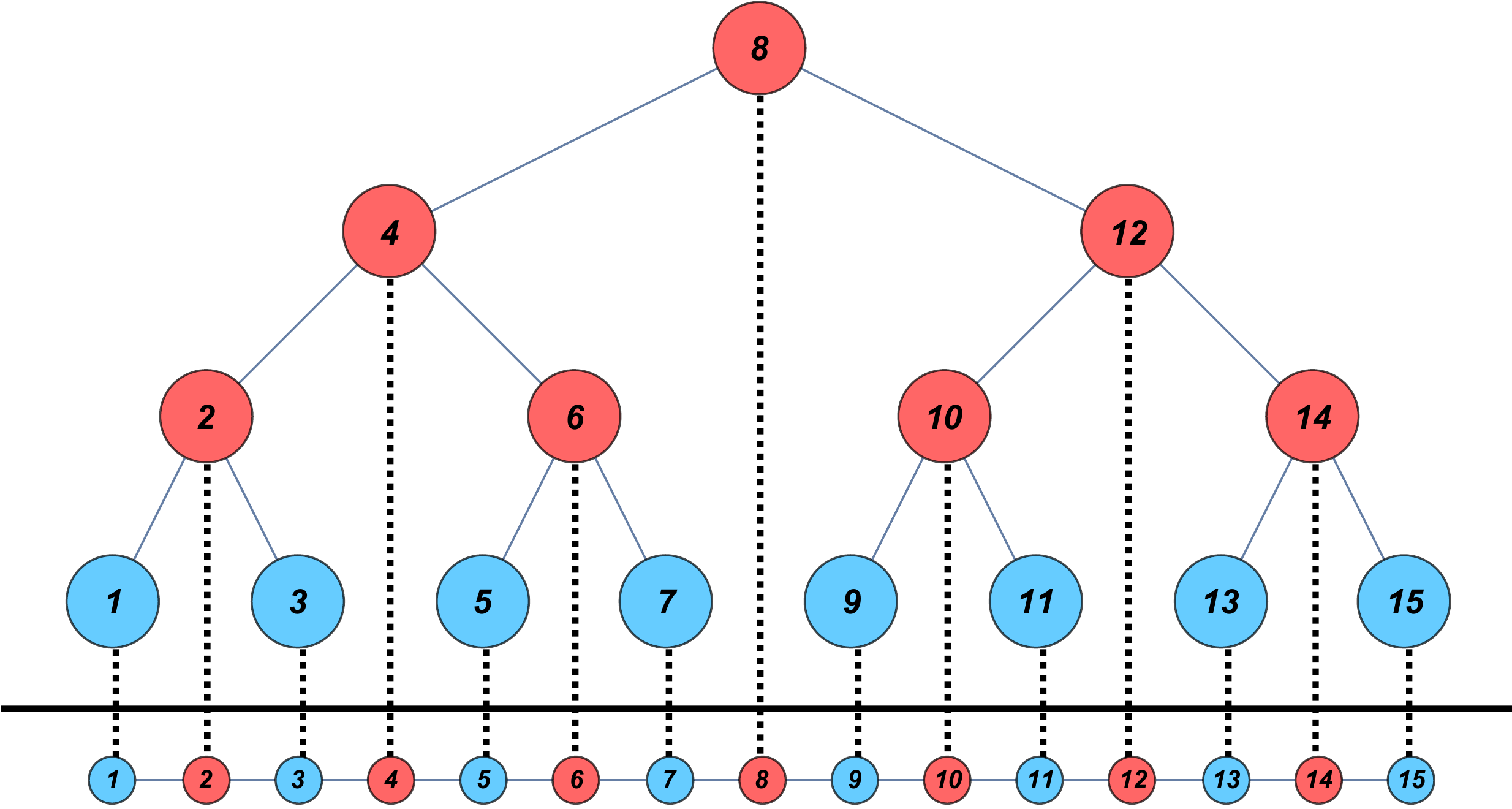}
\caption{\small{Inorder transversal linearization of a binary tree. The color code of the nodes reflects their logical roles: data qubits are colored blue,  while ancilla qubits are colored red.}}
\label{fig:linearized-binary-tree}
\end{center}
\end{figure}  

The schedule for a linear connectivity is obtained by linearizing the binary tree with in-order transversal indexing, as visualized in Fig.~\ref{fig:linearized-binary-tree}. We consider the case $n=2^k$ for convenience, but the algorithm can be easily adapted beyond this condition by considering $k=\lceil\log_2(n)\rceil$. Data qubit $x_i$, with $i=1,2,\dots,n$, is mapped to physical qubit $q_{2i-1}$. The ancilla qubits forming the level immediately above in the binary tree (i.e. having height 1) correspond to physical qubits $q_{4j-2}$ for $j=1,2,\dots,n/2$. Ancilla qubits at greater heights are similarly uniformly distributed in the physical qubit chain.

\begin{algorithm}[H]
	\caption{Linear Grover Diffusion Operation}
	\label{alg:groverline}
	\begin{algorithmic}[1]
		\item [case corresponding to $n=2^k$]
		\item [number of physical qubits is $(2n-1)$]
		\item [logical qubit $x_i$ mapped to physical qubit $q_{2i-1}$]
		\State $i \gets 1$
		\While{$i \leq k$}
		\State $m \gets 2^{i-1}$
		\While{$m > 1$}
		\State $j \gets 0$
		\While{$j < 2^{k-i}$} \textbf{Parallel}
		\State $s \gets (1+2j) 2^i$
		\State \text{SWAP}\ $(q_{s-m}, q_{s-m+1})$
		\State \text{SWAP}\ $(q_{s+m}, q_{s+m-1})$
		\State $j \gets j+1$
		\EndWhile
		\State $m \gets m-1$
		\EndWhile
		\State $j \gets 0$
		\While{$j < 2^{k-i}$} \textbf{Parallel}
		\State $s \gets (1+2j) 2^i$
		\State \text{TOFF}\ $(q_{s-1}, q_{s+1}, q_s)$
		\State $j \gets j+1$
		\EndWhile
		\State $i \gets i+1$
		\EndWhile
		\State \text{Z} $q_n$
		\State $i \gets k$
		\While{$i \geq 1$}
		\State $j \gets 0$
		\While{$j < 2^{k-i}$} \textbf{Parallel}
		\State $s \gets (1+2j) 2^i$
		\State \text{TOFF}\ $(q_{s-1}, q_{s+1}, q_s)$
		\State $j \gets j + 1$
		\EndWhile
		\State $m \gets 2$
		\While{$m \leq 2^{i-1}$}
		\State $j \gets 0$
		\While{$j < 2^{k-i}$} \textbf{Parallel}
		\State $s \gets (1+2j) 2^i$
		\State \text{SWAP}\ $(q_{s-m}, q_{s-m+1})$
		\State \text{SWAP}\ $(q_{s+m}, q_{s+m-1})$
		\State $j \gets j+1$
		\EndWhile
		\State $m \gets m+1$
		\EndWhile
		\State $i \gets i-1$
		\EndWhile
	\end{algorithmic}
\end{algorithm}

\begin{figure}
	\centering
	\includegraphics[width=0.98\linewidth]{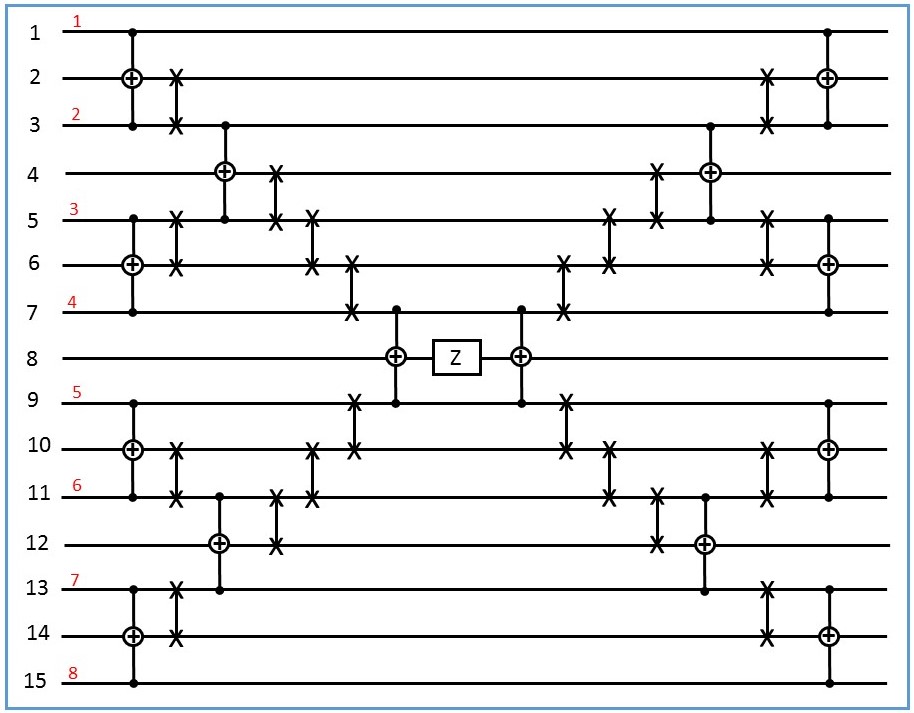}
	\caption{\small{Schedule for the Grover diffusion operator with $n=8$ data qubits, corresponding to a database with $2^8=256$ elements. The implementation also includes $(n-1)=7$ ancilla qubits, those not associated with a red index.}}
	\label{fig:schedule-1d-grover}
\end{figure}

The Toffoli gates at the lowest height of the binary tree can be immediately implemented on the linear configuration, as each pair of control data qubits is adjacent to the target ancilla between them. However, one needs to apply SWAP gates to route the ancilla qubits together to connect the physical qubits that compose the Toffoli gates for the higher levels of the binary tree. This is done by keeping the target qubit of the Toffoli gate fixed, and by moving the two control qubits (one on each side of the target qubit) to the adjacent positions in the linear array of qubits. The pseudocode is presented in Algorithm \ref{alg:groverline} and graphically in Fig.~\ref{fig:schedule-1d-grover}.

\textbf{Performance Analysis}: With an in-order traversal mapping of the binary tree, we can analyze the SWAP overhead between each round of Toffoli gates, and calculate a final performance overhead. In the level at height $h$ of a $k=\log_2{n}$ depth binary tree, where $h$ runs from $1, \cdots, k$, the distance between the ancilla comprising each Toffoli triplet requires $(2^{h}-2)$ SWAPs to be covered\footnote{The depth of a node is the number of edges from the node to the tree's root node. A root node will have a depth of 0. The height of a node is the number of edges on the longest path from the node to a leaf. A leaf node will have a height of 0.}, given by the size of the subtrees that are located between these nodes in an inorder traversal. These can be executed in parallel, as we can SWAP the control qubits both towards the target qubit, which is located between the controls by design. This results in a SWAP chain circuit depth of $2^{\log(n)-h}-1$ for the iteration at height $h$. For an entire binary parity tree then, we add to the parity accumulation circuit a total circuit depth of $\sum_{h=1}^{\log(n)} 2^{\log(n)-h}-1 = n-\log(n)-1$. Adding these to the $\log(n)$ serial Toffoli gate blocks, again each decomposed into a set of 4 single qubit rotations and 3 CNOT operations, results in a full algorithm running time of: $2(3(n-1-\log_2(n))+7\log_2(n))+1 = 6n+8\log_2(n)-5$ time steps, including the execution of the inverted gate sequence and the single qubit gate in the center of the circuit. Crucially, we find that this does not impede the quadratic speedup offered by the algorithm over the classical search counterpart algorithm, given sufficient size of the input space. In fact the overhead is $\mathcal{O}(n)$ while the quantum speedup was $\mathcal{O}(2^{n/2})$.

Two final remarks: first we observe that the initial mapping between logical data qubits and the physical qubits can be relaxed. Since the Grover diffusion operator is a symmetric operation of all data qubits, any initial mapping that alternates a data qubit with an ancilla qubit works equally well. Second, we notice that such configuration can be obtained with at most $n(n-1)/2$ SWAPs and depth of $(n-1)$ sequential SWAPs, the worst initial mapping being all data qubits accumulated at one end of the linear array.

\subsection{Ladder Results}

\subsubsection{Quantum Fourier Transform}
\begin{figure*}
\begin{center}
 \includegraphics[width=\textwidth]{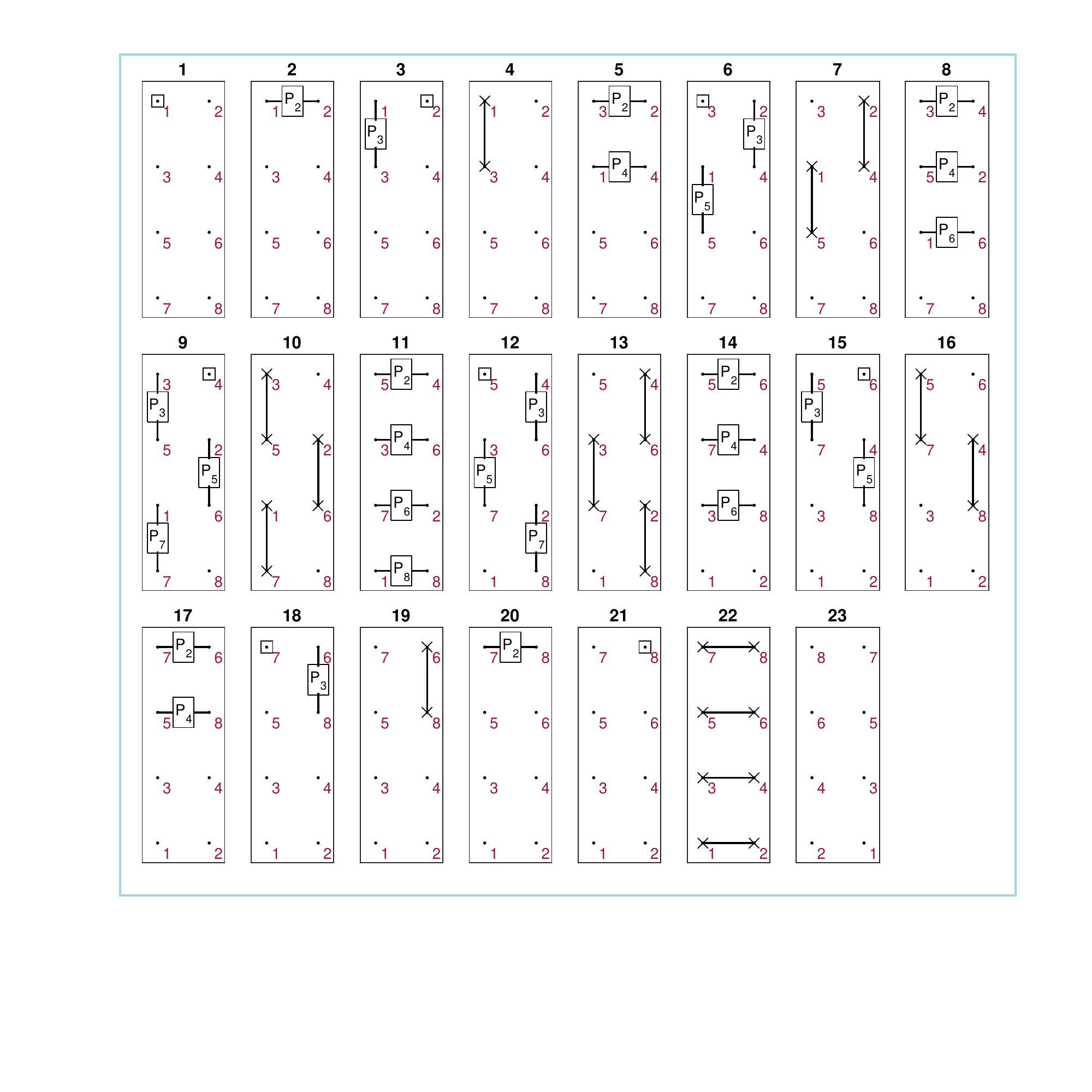}
 \vspace{-100pt}
\caption{\small{Snapshots in time of the QFT being executed on a ladder. The squares represent Hadamard gates. The two-qubit gates used are controlled-phase and SWAP gates. The numbers on top of each sub-figure indicate the temporal order.}}
\label{fig:schedule_qft_ladder}
\end{center}
\end{figure*}

For the QFT, a simple embedding of the linear array onto the ladder of two columns each of height $\frac{n}{2}$ would allow for the application of Algorithm \ref{alg:qftlnn} directly, which would not reduce circuit depth. We improve upon this by designing a schedule that offers a constant-factor speedup over a linear array. We label the physical qubits as shown in Fig.~\ref{fig:hardware}. This labeling scheme will be used to enable direct logical to physical qubit mapping.  For this, we will essentially execute two copies of Algorithm \ref{alg:qftlnn} in parallel, one on each column of the ladder, interspersed with controlled-phase gates between neighboring qubits across the ladder. 
The procedure is presented in Algorithm \ref{alg:ladderqft} and Fig.~\ref{fig:schedule_qft_ladder}.



\begin{algorithm}[H]
	\caption{Ladder Quantum Fourier Transform}
	\label{alg:ladderqft}
	\begin{algorithmic}[1]
    	\State $\text{H } (q_1)$
		\For{$i$ in $\{1,3, 5,\ldots n-3, n, n-2,n-4, \ldots 2\}$}
		    \State $l \gets (i+1)\pmod{2}$
		    \State $j \gets i$\\
		    \textbf{Start Parallel}
		    \While{$j \geq 1$}
		        \State $\text{CP}_{j+l-1}\ (q_{j}, q_{j+(-1)^l})$ 
		        \State $j \gets j-2$
		    \EndWhile\\
		    \textbf{End Parallel}
      	    
      	    \State $j \gets i$\\
      	 \textbf{Start Parallel}
            \While{$j \geq 1$ and $j+2(-1)^l\geq 1$}
                \State $k \gets (j+1)\pmod{2}$
                \State $\text{CP}_{j+2-2l-k}\ (q_{j}, q_{j+2(-1)^l})$
                \State $j \gets j-1-2k$
            \EndWhile
            \State $\text{H} ((j+1)\pmod{2}+1)$\\
        \textbf{End Parallel}
        
        \State $j \gets i$\\
        \textbf{Start Parallel}
            \While{$j\geq 1$ and $j+2(-1)^l\geq 1$}
                \State $k \gets (j+1)\pmod{2}$
                \State $\text{SWAP}(j,j+2(-1)^l)$
                \State $j \gets j-1-2k$
            \EndWhile\\
            \textbf{End Parallel}
		\EndFor
	\end{algorithmic}
\end{algorithm}

\textbf{Performance Analysis}:
This algorithm is composed of $n-1$ iterations of 3 types of parallel blocks. First, we execute a set of ``horizontal'' controlled phase gates that cross the ladder, followed by a set of ``vertical'' controlled phase gates. Lastly we immediately follow the vertical gates with a corresponding set of SWAP operations. The SWAP overhead has been reduced, with respect to the linear case, by only inserting a SWAP timestep once for every two sets of controlled phase operations. The first $n-2$ iterations of the algorithm execute all 3 of these blocks, for an iteration depth of $9$ after canceling the internal CNOTs as was performed for Algorithm \ref{alg:qftlnn}. The last iteration only executes the first of these blocks, adding 4 operations to the critical path, which combines with a final step of SWAP operations and with leading and final Hadamard operations for a total circuit depth of $9n-11$, a small reduction compared to the linear nearest neighbor machine.

\subsubsection{Jordan-Wigner String}
A Jordan-Wigner string on a ladder can be executed in time asymptotically equivalent to that of the linear array with a factor of 2 reduction. The construction proceeds by implementing the PARITY subroutine defined in Algorithm \ref{alg:PARITY} in parallel on each of the two columns of height $n/2$ of the ladder, adding an additional CNOT operation between the center two qubits located in row $\lfloor n/4\rfloor$. The rotation operation is performed on the target of this added CNOT, and the circuit is then reversed. This is described by the pseudocode contained in Algorithm \ref{alg:jwladder}.

\begin{figure}
  \centering
  \includegraphics[width=0.9\linewidth]{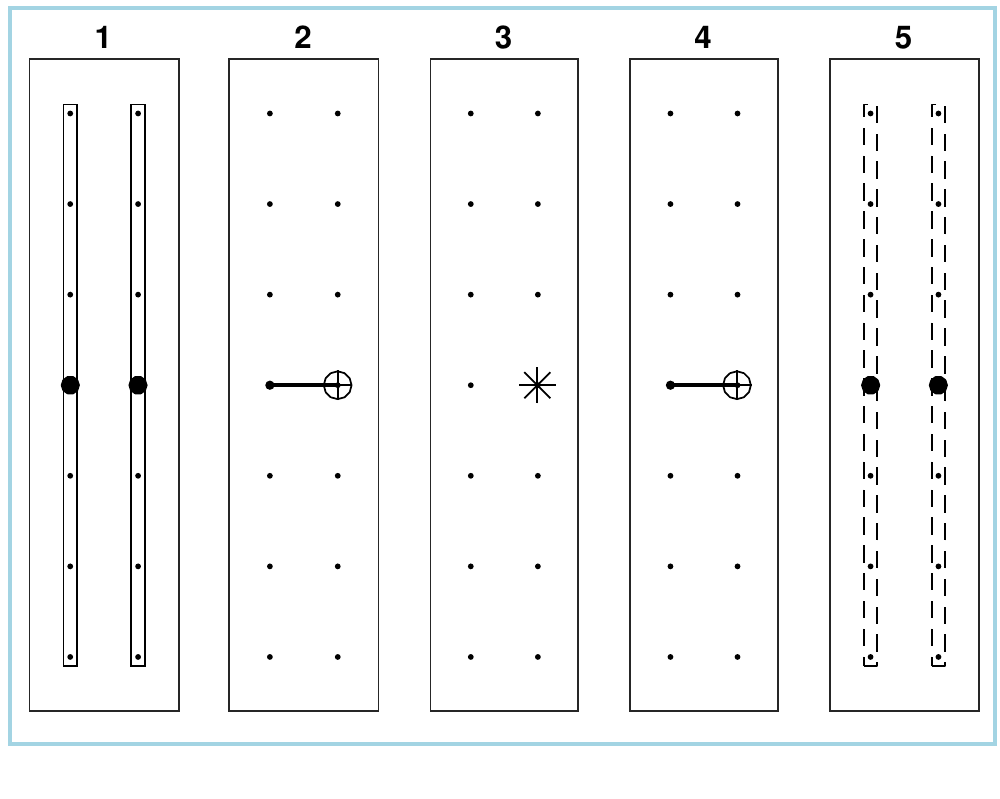}
  \vspace{-10pt}
  \caption{\small{Schedule for the Jordan-Wigner string on a ladder. The solid vertical box is the PARITY operation on a linear array and the dashed vertical box is PARITY$^{\dagger}$. There are CNOT gates between the middle qubits on either leg of the ladder and the $*$ operator indicates a Z-rotation.}}
  \label{fig:schedule-ladder-jw}
\end{figure}

\begin{algorithm}[H]
	\caption{Ladder Jordan-Wigner String: $e^{-i\frac{\theta}{2}\sigma_z^{\tens_n}}$}
	\label{alg:jwladder}
	\begin{algorithmic}[1]
		\State $m \gets \ceil{n/2}$
		\State $k \gets (m+1) \pmod{2}$\\
		\textbf{Start Parallel}
		    \State PARITY(Column 1,$m$)
            \State PARITY(Column 2,$m$)\\
        \textbf{End Parallel}
        
		\State \text{CNOT } ($q_{m-k},q_{m+1-k})$
		\State \text{RZ } ($q_{m+1-k}, \theta$)
		\State \text{CNOT } ($q_{m-k},q_{m+1-k})$\\
		
		\textbf{Start Parallel}
		    \State PARITY$^{\dagger}$(Column 1, $m$)
            \State PARITY$^{\dagger}$(Column 2, $m$)\\
        \textbf{End Parallel}
	\end{algorithmic}
\end{algorithm}

%

\textbf{Performance Analysis}: By splitting the connectivity into two columns each of height $n/2$, an approximate factor of $2$ in circuit depth can be saved by executing Algorithm \ref{alg:jw1sched} on each column in parallel, resulting in a final running time of $n/2+1+(n/2 \mod 2)$, where as before $n$ can be parameterized by $N_{\text{CNOT}}$ and $3N_{\text{SWAP}}$ in order to account for the SWAP gate decomposition. Reported in table \ref{table:depth} is the circuit depth of an instantiation of the algorithm with global parity.

\subsubsection{Grover Diffusion Operator}
To utilize the added connectivity of the ladder, we will consider an extension of the binary parity tree in-order traversal embedding given in section \ref{alg:groverline}. We find that the added connectivity here also allows for approximately a factor of 2 reduction in the total SWAP count and the circuit depth of the resulting circuit. Intuitively, the idea is to use a ladder of two columns each of height $n$ and to extend the linear embedding by first embedding all of the $n$ data qubits into the first column, and using the second column as ancilla. The first operation of the algorithm is to perform the first Toffoli interaction between every pair of input qubits, targeting a single ancilla qubit in the second column. The resulting ancilla qubits in the second column are then spaced out by another yet unused ancilla qubit. This operation requires an additional 2 SWAPs added to the circuit depth of the Toffoli gate in decomposition, to account for the non-adjacency of the one of the control qubits and the target qubit. The remainder of the algorithm implements the linear array algorithm on the second column, which amounts to executing a new binary parity tree subroutine of depth $k'=\log(n/2) = k - 1$. 

\begin{figure}
  \centering
  \includegraphics[width=0.9\linewidth]{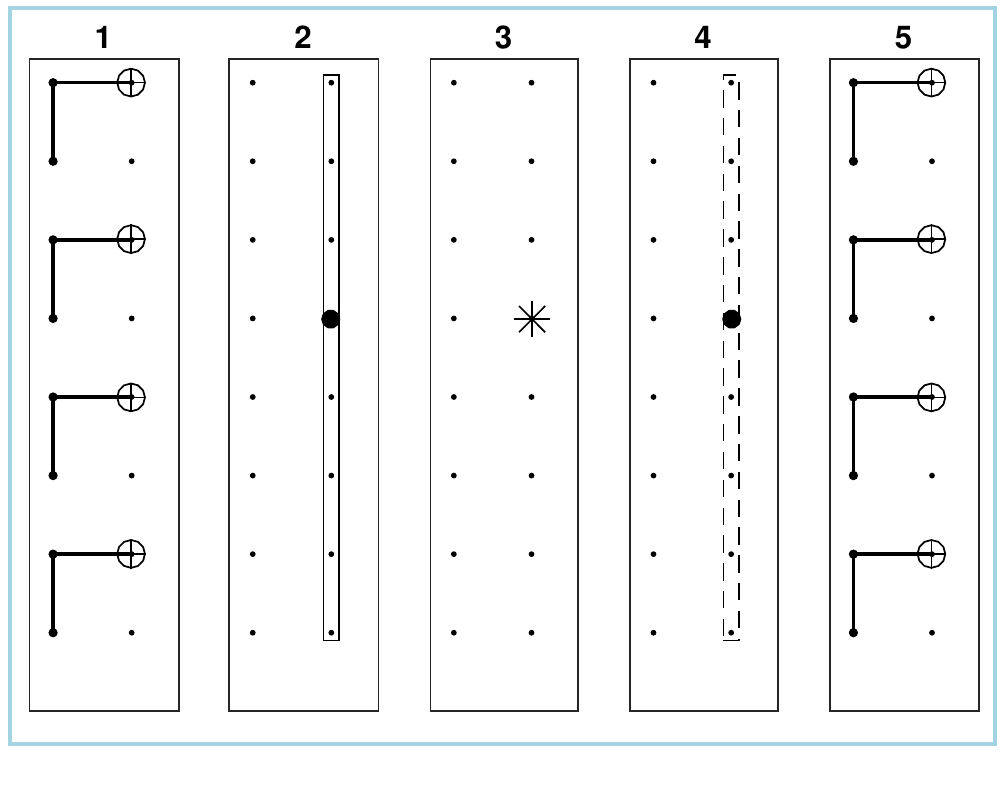}
  \caption{\small{Schedule for the Grover diffusion operation on a ladder. Toffoli gates of steps 1 and 5 are explicitly depicted. Step 3 includes a single Z. The solid box is the AND operation on a line and the dashed box is AND$^{\dagger}$.}}
  \label{fig:schedule-ladder-grover}
\end{figure}

Pseudocode for this construction is specified in Algorithm \ref{alg:groverladder}, noting that the input qubits are positioned in the first column, and the ancilla are all positioned in the second column. The physical indices will be specified again as in Fig.~\ref{fig:hardware}.
\begin{algorithm}[H]
	\caption{Ladder Grover Diffusion Operator}
	\label{alg:groverladder}
	\begin{algorithmic}[1]
		\State $i \gets 1$
		\While{$i < 2n-1$} \textbf{Parallel}
      		\State $\text{TOFF}\ (q_{i},q_{i+2}, q_{i+1})$ 
    		\State $i \gets i + 4$
		\EndWhile
		\State LNNGroverDiffusionOperation(second column)
		\State $i \gets 1$
		\While{$i > 0$} \textbf{Parallel}
      		\State $\text{TOFF}\ (q_{i},q_{i+2}, q_{i+1})$ 
    		\State $i \gets i +4$
		\EndWhile
	\end{algorithmic}
\end{algorithm}

\textbf{Performance Analysis}: The algorithm operates by performing the lowest level of the binary parity tree interaction first, and embedding these results into a linear array fully contained in the second column of the ladder. The remainder of the algorithm is to complete the binary parity tree, which can be completed by directly performing Algorithm \ref{alg:groverline} on this second column.

Notice that, by doing this, the second lowest level of the parity tree becomes adjacent as well. In fact, the performance of this algorithm is equal to the performance of Algorithm \ref{alg:groverline} operating on a set of $n/2$ input qubits, with an additional two Toffoli operation steps. The ladder can perform the full algorithm in circuit depth $6(n/2)+8\log_2(n/2)-5+14+12=3n+8\log_2(n)+13$, reducing the circuit depth compared to the linear array by approximately a factor of 2. 

\subsection{Grid Results}
\subsubsection{Quantum Fourier Transform}
While an implementation of the square grid QFT algorithm can potentially obtain a lower circuit depth than that of the ladder implementation, the performance gap between the ladder and the theoretical lower bound is likely to not exceed the complexity that would be added to build the machine. The magnitude of the performance gap between the ladder and the theoretical lower bound is one of the major results of this work: namely that a limited connectivity device (1D or almost 1D) can perform QFT with near-optimal application latency.

\subsubsection{Jordan-Wigner String}
The 2D-grid can be used effectively to reduce the circuit depth overhead for the Jordan-Wigner string from $\mathcal{O}(n)$ to $\mathcal{O}(\sqrt{n})$ by extending the construction utilized to perform the ladder implementation. 

Specifically, considering a square grid of dimension $\sqrt{n} \times \sqrt{n}$, the transform can proceed by first performing the PARITY subroutine in Algorithm \ref{alg:PARITY} on all columns (rows) of the grid, followed by an invocation of the Linear Array Algorithm \ref{alg:jw1sched} operating on the central row (column). 

\begin{figure}
  \centering
  \includegraphics[width=\linewidth]{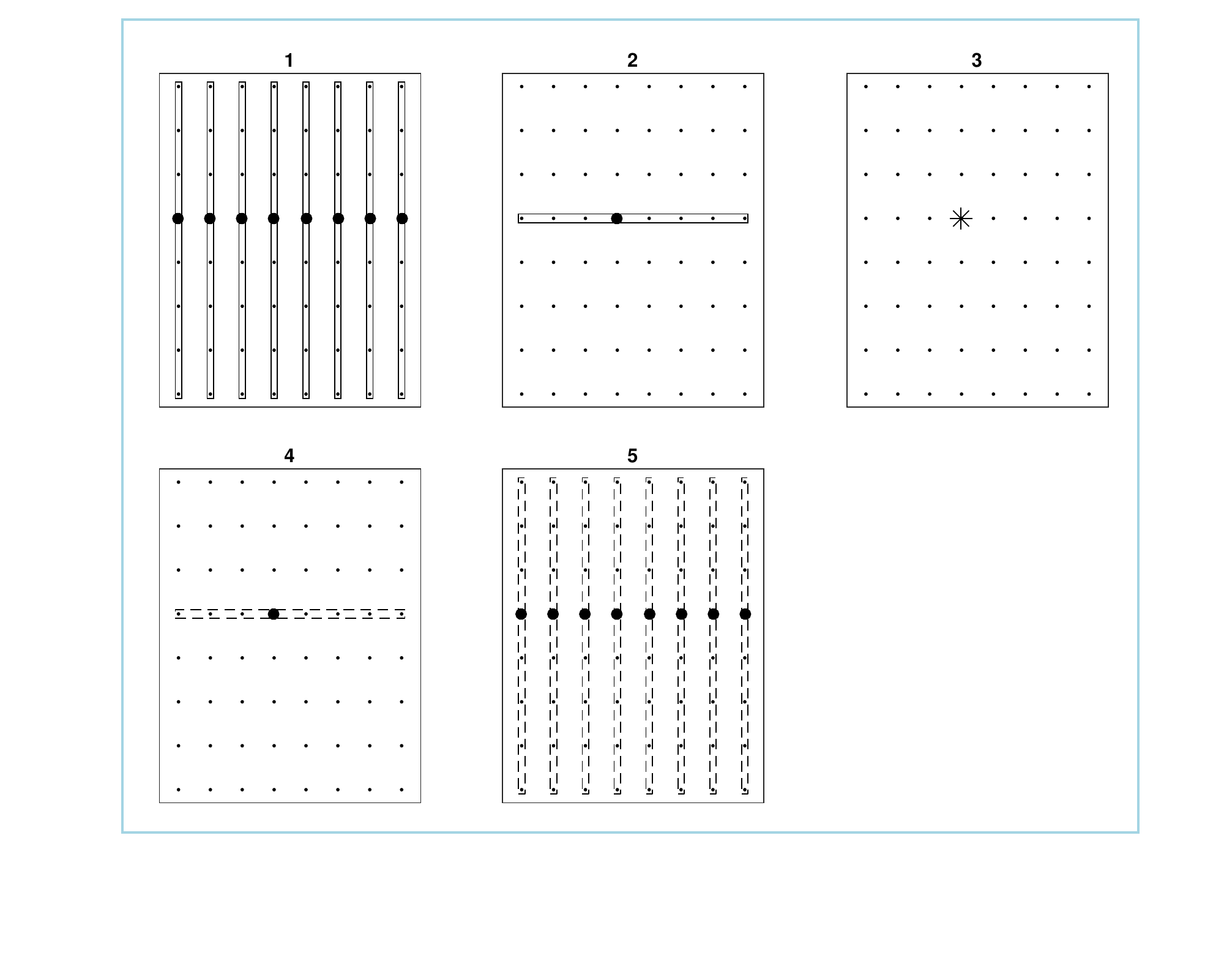}
  \vspace{-30pt}
  \caption{\small{Schedule for the Jordan-Wigner string on a grid. The solid boxes are the PARITY operations on a line and the dashed boxes are PARITY$^{\dagger}$.}}
  \label{fig:schedule-2d-jw}
\end{figure}

\begin{algorithm}[H]
	\caption{Grid Jordan-Wigner String: $e^{-i\frac{\theta}{2}\sigma_z^{\tens_n}}$}
	\label{alg:jwgrid}
	\begin{algorithmic}[1]
		\State m $\gets \sqrt{n}\lfloor(\sqrt{n}-1)/2\rfloor$
		\State $i \gets 1$
		\While{$i \leq \sqrt{n} $} \textbf{Parallel}
			\State PARITY(\text{Column} $i$,$q_{m+i}$)
		\EndWhile
		\State PARITY(\text{Row} $\lceil\sqrt{n}/2\rceil$,$q_{m+\lceil\sqrt{n}/2\rceil}$)
		\State \text{Rz } ($q_{m+\lceil\sqrt{n}/2\rceil}$, $\theta$)
		\State PARITY$^{\dagger}$(\text{Row} $\lceil\sqrt{n}/2\rceil$,$q_{m+\lceil\sqrt{n}/2\rceil}$)
		\State $i \gets 1$
		\While{$i \leq \sqrt{n} $} \textbf{Parallel}
			\State PARITY$^{\dagger}$(\text{Column} $i$,$q_{m+i}$)
		\EndWhile
	\end{algorithmic}
\end{algorithm}

%

\textbf{Performance Analysis}: Altogether, this Algorithm is comprised of four serialized sets of parallel PARITY subroutines, each executed on $\sqrt{n}$ input qubits, as well as the central rotation gate. This combines for a full execution time of $2\sqrt{n}+1+2(\sqrt{n}\mod 2))$, with the appropriate adjustments for the presence of SWAP operations.

\subsubsection{Grover Diffusion Operation}
In order to fully utilize a 2D-grid connectivity for the Grover diffusion operation, we will use both the linear and ladder array implementations as building blocks. Informally, the algorithm will operate on a grid of consisting of $\sqrt{n}$ columns each of length $2\sqrt{n}$, where every column contains $\sqrt{n}$ input qubits and the original $\sqrt{n}-1$ ancilla qubits, and we have inserted an additional row of ancilla qubits into the center of the grid. Every column will perform the AND subroutine, Algorithm \ref{alg:AND}. An additional SWAP is inserted to accommodate the additional row. The results of the AND subroutines will be contained in qubits in row $\ceil{\sqrt{n}}$, at which point Algorithm \ref{alg:groverladder} is executed on the row containing these results and the added ancilla row. This is summarized by the pseudocode of Algorithm \ref{alg:grovergrid}.

Reference \cite{Rosenbaum2012} shows that the optimal depth for 2D architectures and non-adaptive circuits is $\mathcal{O}(\sqrt{n})$  and suggest a constructive procedure in which information is processed from the outer square towards the center of the grid. In our case we obtain the same scaling by using our ladder construction as building block.

\begin{algorithm}[H]
    \caption{AND Subroutine}
    \label{alg:AND}
    \begin{algorithmic}[1]
    \item [input: list of $2n$ qubits indexed in sequence $1, \cdots, 2n$]
    \item [logical qubit $x_i$ mapped to physical qubit $q_{2i-1}$]
	\State $i \gets 1$
	\While{$i \leq k$}
    	\State $m \gets 2^{i-1}$
    	\While{$m > 1$}
    	    \State $j \gets 0$
        	\While{$j < 2^{k-i}$} \textbf{Parallel}
        	    \State $s \gets (1+2j) 2^i$
                \State \text{SWAP}\ $(q_{s-m}, q_{s-m+1})$
                \State \text{SWAP}\ $(q_{s+m}, q_{s+m-1})$
                \State $j \gets j+1$
        	\EndWhile
        	\State $m \gets m-1$
		\EndWhile
	    \State $j \gets 0$
	    \If{$i == k$} 
            \State \text{SWAP}\ $(q_{n-1}, q_{n})$
        \EndIf
    	\While{$j < 2^{k-i}$} \textbf{Parallel}
    	    \State $s \gets (1+2j) 2^i$
            \State \text{TOFF}\ $(q_{s-1}, q_{s+1}, q_s)$
            \State $j \gets j+1$
    	\EndWhile
    	\State $i \gets i+1$
	\EndWhile
    \end{algorithmic}
\end{algorithm}

\begin{algorithm}[H]
	\caption{Grid Grover Diffusion Operation}
	\label{alg:grovergrid}
	\begin{algorithmic}[1]
	\State $k \gets 1$
    \While {$k \leq \sqrt{n}$} \textbf{Parallel}
        \State AND(\text{Column} $k$)
    \EndWhile
    \State LadderGroverDiffusionOperator(center two rows)
    \State $k \gets 1$
    \While {$k \leq \sqrt{n}$} \textbf{Parallel}
        \State AND$^\dagger$(\text{Column} $k$)
    \EndWhile
	\end{algorithmic}
\end{algorithm}

\begin{figure}
  \centering
  \includegraphics[width=\linewidth]{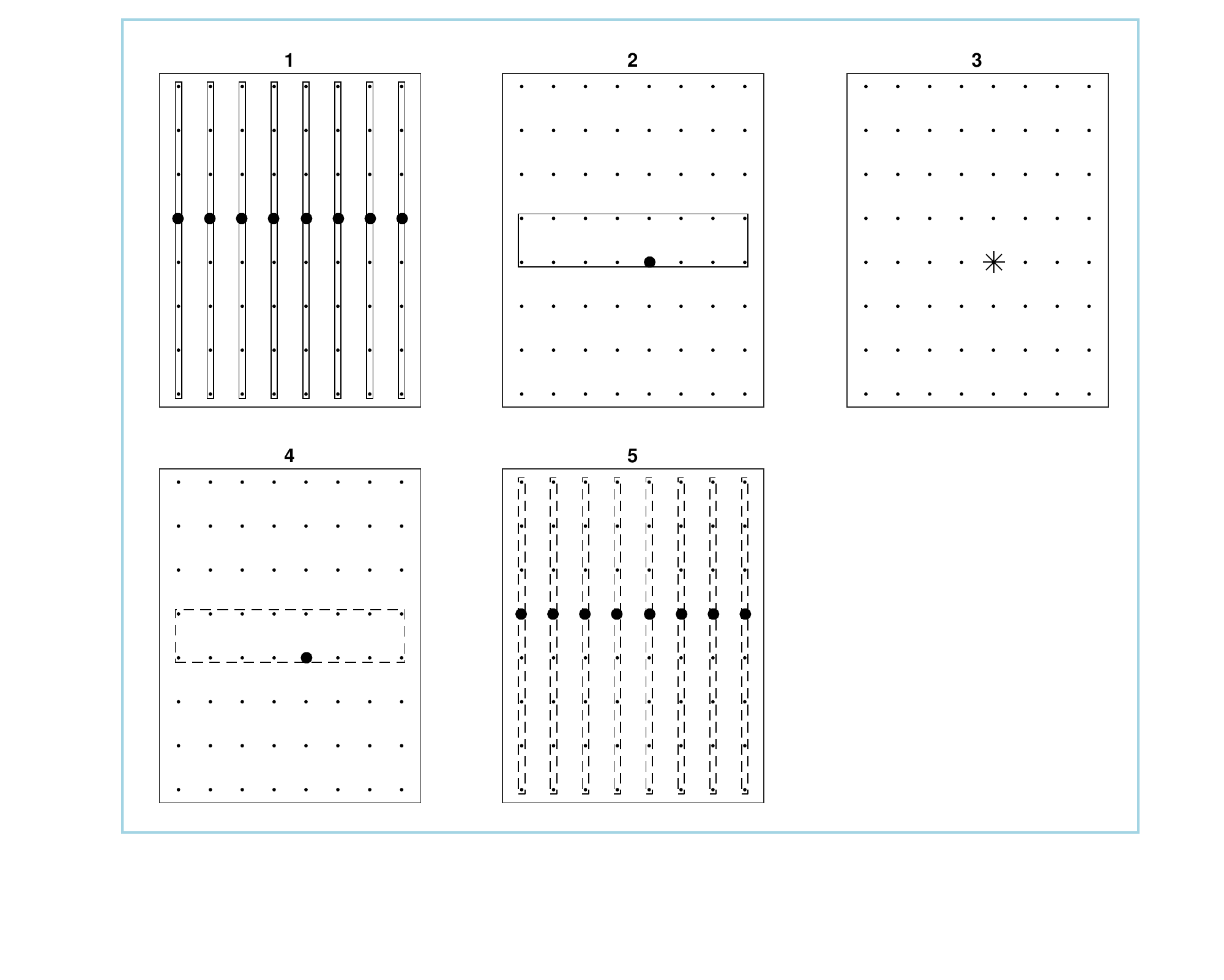}
  \vspace{-30pt}
  \caption{\small{Schedule for the Grover diffusion operation on a grid. The solid boxes are the AND operations on a line and the dashed boxes are AND$^{\dagger}$.}}
  \label{fig:schedule-2d-grover}
\end{figure}

\textbf{Performance Analysis}: The algorithm can execute all of the column operations in parallel, in time $3\sqrt{n}+4\log_2(n)$ given by the circuit depth of the AND subroutine operating on $\sqrt{n}$ input qubits. It will then perform the subsequent Ladder algorithm in another $3\sqrt{n}+8\log_2(\sqrt{n})+13$ time steps. This is followed by another iteration of a set of parallel column AND subroutines, resulting in a final running time of $9\sqrt{n}+8\log_2(n)+13$. This Algorithm makes good use of nearly the entire 2D grid of qubits that are available, and while there are implementations of these multi-controlled NOT operations that do not require ancilla \cite{Barenco}, they add a large expansion to the circuit depth in order to do so. It is likely that an implementation based without added ancilla could save a constant factor of qubits at the expense of a larger expansion in running time. The presented implementation would be therefore minimal in comparison, in terms of the product of the number of qubits and the circuit depth required to execute the algorithm.

%% file: conclusion.tex
This work addresses the important question of quantifying the impact of reduced connectivity of hardware architectures on the performance of quantum algorithms. While many quantum applications have been theoretically proposed without concern for the underlying hardware connectivity, we have shown that for three representative benchmarks, extensive optimization is possible when scheduling to devices of limited connectivity. For the Quantum Fourier Transform in particular, the performance gap between a ladder connectivity graph and that of the all-to-all theoretical lower bound is particularly small, corresponding to an overhead in the constant factor of only around $12.5\%$. However, the other benchmarks we considered show $\mathcal{O}(\sqrt{n})$ performance gaps between ladder and full 2-dimensional grid connectivities, typically related to the maximum distance between any two qubits in the hardware. The primary driver behind these performance figures is scheduling, and as a result it seems that effort spent compiling applications to machines can help reduce the burden on efforts to increase the connectivity of the devices themselves.

While the bounds on the performance impact are relatively small, an increase of a factor of $\sqrt{n}$ may potentially become an impediment to operation of large scale applications in near term devices where the total coherence time available is finite. However, under the safe assumption that large machines will ultimately need to be operated in a fault-tolerant manner, the extra $\sqrt{n}$ time complexity would only impact the actual execution time of applications, and does not factor into application fidelity. It is reasonable to assume that users of such machines will not be significantly affected by an extension of running time of this magnitude, which is at most linear in the number of qubits and logarithmic when compared to both the quantum speedup and the computational complexity of the corresponding classical algorithms.

Our work extends to the scheduling and execution of fully error corrected circuits operating on encoded qubits. In larger machines, while the underlying physical qubit connectivity may be required to be quite dense to support error correcting codes \cite{dennis2002topological,bombin2014structure}, many proposals for large scale architectures rely upon some form of clustering or modularity of logical qubits \cite{brown2016co,jones2012layered}. These scheduling results can motivate these approaches, showing that sparse connectivity between encoded qubit modules may suffice for attaining quantum speedup.

Future work in this direction will involve taking into account control constraints, for which, in addition to limited connectivity, the parallel execution of quantum gates is subjected to additional restrictions due, for example, to the electronics.

%% file: gate-decompositions.tex
\label{sec:gate-decomposition-cp}
For normalization of analysis across the separate applications, we perform analysis with a gate set comprised of single qubit rotations, single qubit phase rotations, and the two-qubit CNOT operation. As a result, each of the gates in the considered benchmarks needs to be decomposed into this basis. Provided here is the explicit form for the \textit{controlled phase gate}. The \textit{Margolus Toffoli gate} can be found specifically in \cite{Barenco}, section VI. B..
\subsection{Controlled Phase Gate C-P$_n$}
We wish to decompose the operation given by equation \ref{eq:cphase} into a sequence of single qubit gates and CNOT operations. This decomposition is provided in the bottom panel of Figure~\ref{fig:qftcirc}, and can be explicitly verified in matrix form as follows:
\begin{align*}
	\text{CP}_{n}(q_1,q_{2}) &= P_{n+1} \otimes P_{n+1} \times \text{CNOT} \times \mathbb{I}_2 \otimes P_{n+1}^{\dagger} \times \text{CNOT}\\ &= {\begin{pmatrix} 1&0&0&0\\0&e^{\frac{2\pi i}{2^{n+1}}}&0&0\\0&0&e^{\frac{2\pi i}{2^{n+1}}}&0\\0&0&0&e^{\frac{2\pi i}{2^{n}}}\end{pmatrix} \times \begin{pmatrix} 1&0&0&0\\0&1&0&0\\0&0&0&1\\0&0&1&0\end{pmatrix} \times \begin{pmatrix} 1&0&0&0\\0&e^{\frac{2\pi i}{2^{n+1}}}&0&0\\0&0&1&0\\0&0&0&e^{\frac{2\pi i}{2^{n+1}}}\end{pmatrix} \times \begin{pmatrix} 1&0&0&0\\0&1&0&0\\0&0&0&1\\0&0&1&0\end{pmatrix}}\\
	&= {\begin{pmatrix} 1&0&0&0\\0&e^{\frac{2\pi i}{2^{n+1}}}&0&0\\0&0&e^{\frac{2\pi i}{2^{n+1}}}&0\\0&0&0&e^{\frac{2\pi i}{2^{n}}}\end{pmatrix} \times \begin{pmatrix} 1&0&0&0\\0&e^{\frac{2\pi i}{2^{n+1}}}&0&0\\0&0&e^{\frac{2\pi i}{2^{n+1}}}&0\\0&0&0&1\end{pmatrix}} =\begin{pmatrix} 1&0&0&0\\0&1&0&0\\0&0&1&0\\0&0&0&e^{\frac{2\pi i}{2^n}}\end{pmatrix} \, ,
\end{align*}
where we have used $P_k = {\begin{pmatrix} 1&0\\0&e^{\frac{2\pi i}{2^k}} \end{pmatrix}}$, and $\mathbb{I}_k$ as the identity matrix of dimension $k$. The CNOT gates act on qubits $q1$ and $q2$ with the former as the control.

\subsection{Margolus Toffoli Gate}\label{sec:gate-decomposition-toffoli}
Following \cite{Barenco} we make use of the Toffoli gate decomposition as four single qubit gates and three two qubit gates, up to phase incurred on nonzero states. This phase accumulation does not have a functional effect in the Grover Diffusion circuitry we consider in this work, which enables the use of this optimized decomposition. Specifically, the Toffoli gate can be written as:
\begin{equation*}
    TOFF(q_1, q_2, q_3) = R(q_3) \times CNOT (q_2, q_3) \times R(q_3) \times CNOT (q_1, q_3) \times R^{\dagger}(q_3) \times CNOT (q_2, q_3) \times R^{\dagger}(q_3)
\end{equation*}
Where $R$ is a rotation about the $Y$ axis of $\pi/8$.

%% file: main.bbl
\begin{thebibliography}{39}
\bibitem [{jwa()}]{jwapp}%
\BibitemOpen
\href@noop {} {\enquote {\bibinfo {title} {Results are presented for
			instantiations of the application that do not include gaps -- i.e. all qubits
			are involved in the parity calculation},}\ }\BibitemShut {NoStop}%
\end{thebibliography}
